\documentclass[%
 reprint,
 superscriptaddress,
 amsmath,amssymb,
 aps, 
]{revtex4-2}

\usepackage{graphicx}
\usepackage{dcolumn}
\usepackage{bm}

\usepackage{multirow}
\usepackage{bm}
\usepackage{textcomp}
\usepackage{longtable}
\usepackage{booktabs}
\usepackage{epsfig}
\usepackage{rotating}
\usepackage{xfrac}
\usepackage{threeparttable}
\usepackage{lipsum} 
\usepackage{float}
\usepackage{dcolumn}
\usepackage{multirow}

\usepackage{rotating}
\RequirePackage{hyperref}
\hypersetup
{
unicode=false,%
pdftoolbar=true,
pdfmenubar=true,
pdffitwindow=false,
pdfstartview=FitH,
pdftitle={PRC},
pdfauthor={LAM Yi Hua},
pdfsubject={Nuclear Physics},
pdfcreator={LAM Yi Hua},
pdfproducer={LAM Yi Hua},
pdfkeywords={NucPhys},
bookmarksnumbered,%
colorlinks=true,%
linkcolor=magenta,%
citecolor=blue,
filecolor=cyan,%
urlcolor=blue,
breaklinks, 
plainpages=false}
\usepackage{bibunits}
\defaultbibliographystyle{apsrev4-1_titled}
\defaultbibliography{my-final-bib-from-jabref_titles,references}

\usepackage{CJKutf8}
\newcommand{\cntextTraKai}[1]{\begin{CJK*}{UTF8}{bkai}#1\end{CJK*}}

\usepackage{xcolor}
\definecolor{Mycolor}{HTML}{C5E0B3}
\definecolor{red}{HTML}{ee6677}
\definecolor{red2}{HTML}{bb5566}
\definecolor{red3}{HTML}{cc3311}
\definecolor{magenta}{HTML}{ee3377}
\definecolor{purple}{HTML}{aa3377}
\definecolor{orange}{HTML}{ee7733}
\definecolor{deepyellow}{HTML}{ddaa33}
\definecolor{yellow}{HTML}{ccbb44}
\definecolor{blue}{HTML}{4477aa}
\definecolor{darkblue}{HTML}{004488}
\definecolor{teal}{HTML}{009988}
\definecolor{green}{HTML}{228833}
\definecolor{cyan}{HTML}{66ccee}
\definecolor{grey}{HTML}{bbbbbb}

\begin{document}

\preprint{APS/123-QED}

\title{Nuclear Mass Models with Deformation Optimized and Diagnosed by\\Bayesian Analysis with Markov Chain Monte Carlo}%

\author{Xiangnan~Lee~(\cntextTraKai{李祥楠})}%
  \affiliation{Zhejiang Key Laboratory of Quantum State Control and Optical Field Manipulation, \\Department of Physics, \href{https://ror.org/03893we55}{Zhejiang Sci-Tech University}, Hangzhou 310018, China}

\author{Yi~Hua~Lam~(\cntextTraKai{藍乙華})}%
  \email{lamyihua@zstu.edu.cn}
  \homepage{https://lamyihua.github.io}
  \affiliation{Zhejiang Key Laboratory of Quantum State Control and Optical Field Manipulation, \\Department of Physics, \href{https://ror.org/03893we55}{Zhejiang Sci-Tech University}, Hangzhou 310018, China}
  \affiliation{Astrophysical Big Bang Laboratory, Pioneering Research Institute, \href{https://ror.org/01sjwvz98}{RIKEN}, Wako, Saitama 351-0198, Japan}

\author{Zi-Ao~Zhang~(\cntextTraKai{張子奧})}%
  \affiliation{Zhejiang Key Laboratory of Quantum State Control and Optical Field Manipulation, \\Department of Physics, \href{https://ror.org/03893we55}{Zhejiang Sci-Tech University}, Hangzhou 310018, China}

\author{Jayke~Ren~(\cntextTraKai{任可})}%
  \affiliation{Zhejiang Key Laboratory of Quantum State Control and Optical Field Manipulation, \\Department of Physics, \href{https://ror.org/03893we55}{Zhejiang Sci-Tech University}, Hangzhou 310018, China}

\date{\today}%

\begin{abstract}
We employ a full Bayesian analysis with adaptive Metropolis-Hastings Markov chain Monte Carlo (BA-MCMC) sampling to systematically study the posterior probability distributions of the strengths of energy terms in optimized nuclear mass models of Bethe-Weizs\"{a}cker variants. Strong correlations of some energy terms for some mass models are revealed through the parameter degeneracy diagnosis. We analyze selected refined models to determine parameter degeneracies while proposing a new macroscopic-microscopic mass model, BWL, which considers quadrupole and high-multipole deformation and shell corrections. All mass models in this work are analyzed and optimized through the BA-MCMC method. Compared with 2242 precise experimental binding energies of AME2020, BWL produces a root-mean-square deviation of 759~keV, particularly improving the description of masses in the light-nuclei and actinide regions. BA-MCMC offers robust inference on parameter degeneracy while providing an optimization method for future nuclear mass models.
\end{abstract}

\maketitle


\section{\label{sec:intro}Introduction}

The Bethe-Weizs{\"a}cker semi-empirical mass formula \cite{Weizsacker1935, Bethe1936} expresses the liquid drop model (LDM) {\it first} suggested by \citet{Gamow1930}. After systematic refinement by \citet{Bethe1936}, early pioneers such as \citet{Meitner1939} used LDM to explain newly discovered fission by \citet{Hahn1939}, and later the community used the mass model to predict a wide range of nuclear binding energies beyond the reach of contemporary experiments. The Bethe-Weizs{\"a}cker (BW) mass model assuming the nucleus as a very high density drop of incompressible fluid, with its nucleons held together by the nuclear force. The BW model is implicitly characterized by several physical terms, namely, volume, surface, Coulomb, and symmetry energies, and has become one of the classic phenomenological models despite not being able to describe the binding energies at the regions of historical magic numbers. The strengths of these physical terms are kept up to date with least-squares fit (and refinements) using contemporary experimental data, e.g., \citet{Evans1955}, \citet{Gunter1959}, \citet{Bauer1965}, \citet{Myers1976}, \citet{Tseng1980}, \citet{Eisberg1985}, \citet{Rohlf1994}, \citet{Samanta2002}, \citet{Kirson2008}, \citet{Mavrodiev2019}, \citet{Gjorgievska2024}, and \citet{Wu2025}. Numerous further remarkable refinements were made since the work of Bethe and Bacher, for instance, \citet{Spanier1988} added the empirical deformation and shell corrections to the BW model, whereas \citet{Kirson2008} refined the model by introducing several new terms, i.e., Coulomb exchange, Wigner, pairing, surface symmetry, and curvature energies, and also shell correction. More recently, \citet{Wu2025} adopted the isospin-dependent symmetry energy term with surface diffuseness of the WS4 model~\cite{Wang2014}, reconstructed the symmetry energy term, and introduced regional shell corrections. They then formed the BWN mass model, which significantly reduces the discrepancy between experimental and theoretical binding energies, achieving a root-mean-square deviation of 887~keV. Although all these refinements improve the original BW mass model, these new mass models of BW variants still exhibit noticeably systematic deviations in strongly deformed regions, especially in the rare-earth region.

Recent remarkable advances in experimental techniques lay the foundation for the establishment of state-of-the-art radioactive isotope beam facilities, opening unprecedented opportunities to synthesize very short-lived nuclei far from the valley of stability \cite{Yamaguchi2021, Nan2025}, e.g., the Facility for Rare Isotope Beams (FRIB) in the US \cite{Thoennessen2010}, the Radioactive Ion Beam Factory (RIBF) at RIKEN in Japan \cite{Motobayashi2010}, the Second Generation System On-Line Production of Radioactive Ions (SPIRAL2) at GANIL in France \cite{Gales2010}, the recently commissioning High-Intensity Heavy-Ion Accelerator Facility (HIAF) in China \cite{Zhou2022}, and the currently under construction Facility for Antiproton and Ion Research (FAIR) in Germany \cite{Sturm2010} and the Rare Isotope Accelerator complex for Online Experiments (RAON) in Korea \cite{Tshoo2013}.

Nevertheless, many more exotic nuclei of great interest in nuclear astrophysics are still beyond the reach of these facilities. Hence, robust and cost-effective nuclear mass models capable of precisely predicting binding energies across a wide range of nuclei are needed, offering (i) prior information for experimental measurements \cite{Lunney2003}, (ii) the essential base for constructing macroscopic-microscopic approaches, which implicitly take into account quantum shell effects, e.g., the WS3 and WS4 models \cite{Wang2014, Wang2013WS3}, (iv) the high precision training database for developing artificial intelligence (AI) mass models \cite{Mumpower2022, Niu2022, Lu2026, Munoz2025}, (v) interpretable physics features for physics-informed AI approaches \cite{Yuksel2024, Bentley2025, Huang2025}, (vi) the determination of proton drip line important for studying rapid-proton capture process \cite{Woosley2004, Schatz2006, Parikh2008, Parikh2009, Cyburt2016, Schatz2017, Lam2022a, Lam2022b, Lu2024, Lam2025}, (vii) single- and two-neutron separation energies indispensable for (n,$\gamma$) and (2n,$\gamma$) reaction rates for investigating rapid neutron-capture process \cite{Mumpower2015b, Mumpower2016, Marketin2016, Niu2019b, Ma2019, Shi2021, Cowan2021, Hao2023}, and (viii) a forefront reference for mass compilations using microphysics approaches, e.g., binding energies based on the relativistic Hartree-Bogoliubov approach \cite{Liu2024} with PC-X \cite{Taninah2020}, PC-L3R \cite{Liu2023}, DD-PCX \cite{Yuksel2019}, and DD-MEX \cite{Taninah2020} interactions and relativistic Hartree-Bogoliubov approach in continuum \cite{Zhang2022,Guo2024} with PC-PK1 interaction \cite{Zhao2010}.

The construction of robust and cost-effective mass models involves the procedure of refining the classical BW model either by introducing new terms on top of the original ones and modifying these new terms or by refitting the strengths (or parameters) of the original terms based on new experimental inputs. All of these procedures use the least-squares fit to optimize the parameters. Although the least-squares fit method can be applied to estimate parameters of nonlinear models, it inherently relies on local linear approximations and can only provide point estimates and asymptotic standard errors \cite{Jennrich1969,BatesWatts1988}. Moreover, it does not accurately capture the actual uncertainty of the parameters. The approximated errors could be substantial, especially if the sample size is limited or the model is highly nonlinear. In fact, the classical BW model and its variants are formulated as a nonlinear function calibrated with parameters and coefficients dependent on the mass number and proton number. The parameter space exhibits strong nonlinearity and multi-scale features, resulting in a complex mapping between the binding energy, mass and proton numbers.

{\it Bayesian inference} offers an alternative to exploring the parameter space. By treating unknown parameters as random variables and combining prior information with the likelihood of observed data, it gives the full posterior distribution that quantifies all sources of uncertainty. This posterior distribution directly reflects the nonlinear structure of the model without requiring linearization and provides a complete description of parameter uncertainty, including correlations among parameters \cite{Kejzlar2020,Saito2024,Zhang2024NBMA,Phillips2021,Kejzlar2020,Xu2021,Piarulli2023}.

For the mass models of BW variants, the parameters are physically correlated. For instance, the volume and surface coefficients may exhibit a compensatory effect when fitting with nuclear masses. Least-squares fit method can only provide point estimates and marginal confidence intervals, and even if a covariance matrix is computed, it is based on local linear approximations (e.g., the inverse of the Fisher information matrix and maximum likelihood estimation), which could be distorted in nonlinear models \cite{Hauschild2001,Toivanen2008,Mendoza-Temis2008,Xie2024,Bertsch2005}. For this scenario, the Bayesian posterior distribution naturally encodes the joint parameter dependencies. By studying the posterior, one can obtain samples that reveal the correlation structure, enabling accurate estimation of correlations between any two parameters, including nonlinear dependencies. This is crucial for understanding the physical meaning of the model, diagnosing parameter degeneracy, and performing subsequent uncertainty propagation \cite{Kejzlar2020,Saito2024,Zhang2024NBMA,Phillips2021,Kejzlar2020,Xu2021,Piarulli2023}.

{\it Bayesian analysis with the Markov chain Monte Carlo sampling method.}
In the Bayesian framework, when the model is complex, the parameter dimension is high, and the likelihood function has a complicated form, the posterior distribution could not be obtained directly through analytical methods. The Markov chain Monte Carlo (MCMC) sampling method generates approximately independent samples from the target posterior distribution to support subsequent statistical inference. Its advantages include  
(i) handling complex and high-dimensional posterior distributions \cite{Kejzlar2020,Saito2024,Phillips2021,Roberts1997}, 
(ii) avoiding linear approximations and thus preventing the introduction of unknown errors \cite{Saito2024,Lundquist2020,Kejzlar2020}, 
(iii) fully characterizing uncertainty \cite{Saito2024,Phillips2021,Kejzlar2020}, 
(iv) revealing parameter correlation structures through posterior distributions \cite{Kejzlar2020,Saito2024,Lundquist2020}, and 
(v) being applicable to various complex models, e.g., hierarchical models and state-space models \cite{Drischler2024,Phillips2021,Lundquist2020}.

A critical yet overlooked limitation in available nuclear mass uncertainty quantification is the use of linear correlation diagnosis to analyze the inherently non-linear parameter landscapes. For instance, the recent work by \citet{Xu2026} assumed that the collinearity of a 12-parameter LDM could be analyzed using the Pearson correlation coefficients, generating a large condition number and causing them to conclude that the parameter space was not properly constrained and then to implement L2 regularization to constrain the coefficients. In fact, such a large condition number is an artifact of the presumed linearity in the parameter space. For mass models of BW variants, parameter compensations manifest the non-linear characteristics; however, the Pearson correlation coefficient is only suitable for presenting the degree of linear correlation of parameters. In this case, using the Spearman rank correlation matrix (rank transform method) \cite{Bertolli2013,Stephanou2021,Mohan2024} is more appropriate for analyzing non-linear correlations among parameters of a mass model.

{\it Nuclear mass model with deformation.}
Almost all previous mass models of BW variants do not take into account the essential nuclear deformation, i.e., the nuclei are assumed to be spherical, for instance, the model of \citet{Kirson2008} consists of several higher-order terms but does not take into account the contribution of equilibrium deformation to the binding energy, whereas the BWN model proposed by \citet{Wu2025} significantly improves the symmetry energy and shell correction terms, but its macroscopic part is only based on the spherical LDM without deformation-dependent factors. Nuclear deformation is not systematically introduced into the LDM macroscopic energy of both models. As a result, these models show (large) systematic deviations in strongly deformed regions, especially in the rare-earth and actinide regions. \citet{Spanier1988} refined the BW mass model by introducing deformation corrections for each region between magic numbers. Nevertheless, these corrections are introduced to the entire set of nuclei of each region with only specific phenomenological deformation parameters (Table~A of Ref.~\cite{Spanier1988}).

\newpage
In this work, we construct a framework of full Bayesian analysis with adaptive Metropolis-Hastings Markov chain Monte Carlo (BA-MCMC) sampling to analyze and optimize selected mass models of BW variants. Meanwhile, we also propose a new mass model, implicitly taking into account the deformation specifically for each isotope, and then analyze the new model using BA-MCMC and Spearman rank correlation matrix. The paper is organized as follows. We briefly present the nuclear mass models in Sec.~\ref{sec:mass_models}, the inclusion of deformation in Sec.~\ref{sec:deformation}, followed by the implementation of BA-MCMC in Sec.~\ref{sec:BA_MCMC}. Then, we discuss the analysis of BA-MCMC on mass models in Sec.~\ref{sec:BW_MCMC}, \ref{sec:BWK_MCMC}, \ref{sec:BWN_MCMC}, \ref{sec:BWL_MCMC}, and the general improvement with the introduction of deformation is given in \ref{sec:general_improvement}. A discussion of the advantage of BA-MCMC is presented in Sec.~\ref{sec:advantage}. We then summarize our findings in Sec.~\ref{sec:summary}.

\section{Method}
\label{sec:method}

\subsection{Nuclear mass models}
\label{sec:mass_models}

The historical Bethe-Weizs{\"a}cker semi-empirical mass formula consists of the volume energy, surface energy, Coulomb energy, and symmetry energy \cite{Weizsacker1935,Bethe1936}, 
\begin{equation}
B = \alpha_\mathrm{v} A + \alpha_\mathrm{s} A^{2/3} + \alpha_\mathrm{c} Z^2 A^{-1/3} + \alpha_\mathrm{sym} (N - Z)^2 A^{-1} \, ,
\end{equation}
where $B$ is the binding energy of an atomic nucleus with the proton number, $Z$, the neutron number, $N$ and the mass number, $A$. 
Kirson improved the Bethe-Weizs{\"a}cker formula by adding the Coulomb exchange energy, $\alpha_\mathrm{xc} Z^{4/3} A^{-1/3}$, Wigner energy, $\alpha_\mathrm{w} |N - Z| A^{-1}$, pairing energy, $\delta \alpha_\mathrm{p} A^{-1/2}$, surface symmetry energy, $\alpha_\mathrm{st} (N - Z)^2 A^{-4/3}$, curvature energy, $\alpha_\mathrm{r} A^{1/3}$, and shell correction, $\alpha_\mathrm{m} P + \beta_\mathrm{m} P^2$ \cite{Kirson2008}, 
\begin{equation} 
\begin{split}
B = &~ \alpha_\mathrm{v} A + \alpha_\mathrm{s} A^{2/3} + \alpha_\mathrm{c} Z^2 A^{-1/3} + \alpha_\mathrm{sym} (N - Z)^2 A^{-1}\\
& + \alpha_\mathrm{xc} Z^{4/3} A^{-1/3} + \alpha_\mathrm{w} |N - Z| A^{-1} + \delta \alpha_\mathrm{p} A^{-1/2}\\
& + \alpha_\mathrm{st} (N - Z)^2 A^{-4/3} + \alpha_\mathrm{r} A^{1/3}\\
& + \alpha_\mathrm{m} P + \beta_\mathrm{m} P^2 \ ,
\end{split}
\end{equation}
where $\delta = \{(-1)^Z + (-1)^N\}/2$ and $P =\nu_\mathrm{p} \nu_\mathrm{n}/(\nu_\mathrm{p} + \nu_\mathrm{n})$, $\nu_\mathrm{p}$ and $\nu_\mathrm{n}$ are the proton and neutron valence numbers, respectively. The proton (neutron) valence number is obtained by subtracting the nearest proton (neutron) magic number from the proton (neutron) number. The proton magic numbers are 2, 8, 20, 28, 50, 82, 126, whereas the neutron magic numbers are 2, 8, 20, 28, 50, 82, 126, and 184. Hereinafter, the mass model improved by Kirson is labeled as BWK.

Using Kirson's work \cite{Kirson2008} and the WS4 mass model \cite{Wang2014}, \citet{Wu2025} fold the symmetry energy, surface symmetry, and Wigner terms by introducing $\alpha_{\mathrm{sym},I} I^2 A f_s$, where $f_\mathrm{s}$ is the surface diffuseness. They redefine the pairing term as $\delta_\mathrm{np} \alpha_\mathrm{p} A^{-1/3}$, add a linear term of valence nucleons, $c_\mathrm{m} (v_n + v_p)$, and an exponential term, $e_{m1} \delta_\mathrm{shell} e^{e_{m2} (v_p^2 + v_n^2)}$, with $\delta_\mathrm{shell}$ to further refine the description of shell effects. Hereinafter, the mass model constructed by \citet{Wu2025} is retained as BWN, following the designation introduced by them, and it reads 
\begin{equation}
\begin{split}
\label{eq:BWN}
B = &~ \alpha_\mathrm{v} A + \alpha_\mathrm{s} A^{2/3} + \alpha_\mathrm{c} Z^2 A^{-1/3} + \alpha_{\mathrm{sym},I} I^2 A f_s \\
&+ \delta_{np} \alpha_\mathrm{p} A^{-1/3} + \alpha_{xc} Z^{4/3} A^{-1/3} + \alpha_\mathrm{r} A^{1/3} \\
&+ \alpha_\mathrm{m} P + \beta_\mathrm{m} P^2 + c_\mathrm{m} (v_n + v_p) \\
& + e_\mathrm{m1} \delta_{\mathrm{shell}} e^{e_\mathrm{m2} (v_\mathrm{p}^2 + v_\mathrm{n}^2)},
\end{split}
\end{equation}
where
\begin{align}
\alpha_{\mathrm{sym},I} &= c_{\mathrm{sym}} \left( 1 - \frac{k}{A^{1/3}} + \xi \frac{2 - |I|}{2 + |I|A} \right),\nonumber \\
I &= \frac{N - Z}{A}, \nonumber \\
f_\mathrm{s} &= 1 + \kappa_\mathrm{s} \left( \left( I - \frac{0.4A}{A+200} \right)^2 - I^4 \right) A^{1/3}, \nonumber \\
\delta_\mathrm{np} &=
\begin{cases}
(2 - |I| - I^2) 17/16, N \text{ and } Z \text{ even}, \nonumber \\
|I| - I^2, N \text{ and } Z \text{ odd}, \nonumber \\
1 - |I|, N \text{ even}, Z \text{ odd}, \text{ and } N > Z, \nonumber \\
1 - |I|, N \text{ odd}, Z \text{ even}, \text{ and } N < Z, \nonumber \\
1, N \text{ even}, Z \text{ odd}, \text{ and } N < Z, \nonumber \\
1, N \text{ odd}, Z \text{ even}, \text{ and } N > Z. \nonumber
\end{cases}
\end{align}
and the piecewise constant, $\delta_{\mathrm{shell}}$, defines the shell correction effect for each region as below, 
\begin{align}
\delta_{\mathrm{shell}} &=
\begin{cases}
-1, & Z, N \in [8, 24], \\
0, & Z \in [8, 24] \ \&\ N \in (24, 66], \\
0, & Z \in (24, 39] \ \&\ N \in [8, 66], \\
1, & \text{elsewhere}.
\end{cases}
\end{align}
By establishing a region-dependent shell correction, the BWN model achieves a major improvement in describing the nuclear masses, particularly regions surrounding the magic numbers. In general, they obtained the root-mean-square deviation value (rms) of 887~keV for the BWN model by comparing the experimental and modeled data. See the rms of BWN in Table~2 of Ref.~\cite{Wu2025}.

\subsection{Deformation and shell correction}
\label{sec:deformation}

We incorporate deformation effects and a phenomenological shell correction into the BWN liquid-drop model within the macroscopic-microscopic framework. The deformation is accounted for by a multiplicative enhancement factor applied to the macroscopic energy, using the quadrupole and high-multipole deformation parameters \(\beta_2\), \(\beta_4\), and \(\beta_6\) taken directly from the WS4 model~\cite{Wang2014}. The shell correction term is constructed phenomenologically to reproduce the extra binding near magic numbers while being strongly quenched at large deformation (shell quenching), a feature commonly observed in microscopic calculations. 

This approach retains the computational efficiency of LDMs for large-scale nuclear surveys, while capturing the essential quantum shell effects and deformation dependence without solving single-particle spectra or applying the full Strutinsky smoothing procedure. A similar macroscopic-microscopic strategy, but employing genuine Strutinsky shell corrections derived from deformed Woods-Saxon single-particle levels, was used by~\citet{Wang2014} to construct the WS4 mass model. The recent physics-informed neural network mass model of~\citet{Huang2025} also follows the macroscopic-microscopic approach, although with different implementations of shell corrections. In fact, within the macroscopic-microscopic framework, there also exist approaches that incorporate deformation without relying on the Strutinsky shell correction. In earlier WS models, the macroscopic energy is computed directly from the Skyrme energy density functional, and deformation effects are included via the extended Thomas-Fermi approximation. By introducing degree-of-freedom of deformation (\(\beta_2\), \(\beta_4\), and \(\beta_6\)) directly into the macroscopic part and using parameterized density distributions, these methods avoid the complexity of single-particle level smoothing while still capturing contributions of deformation to the total energy \cite{Washiyama2024}.

Hence, the deformation and phenomenological shell corrections are incorporated into the LDM energy of a spherical nucleus generated from the BWN model to yield the macroscopic-microscopic binding energy of nucleus \(i\), 
\begin{align}
\label{eq:BWL}
B_i^{\mathrm{Th}} = &~B_i^\mathrm{macro} \prod_{k=2,4,6} \left(1 + b_k\beta_k^2\right) \nonumber \\
&~+ B_i^\mathrm{shell} \exp\left(-\gamma \sum_{k=2,4,6}\beta_k^2 \right) \, ,
\end{align}
\noindent
where the deformation correction factor \cite{Wang2010a}, 
\begin{align}
\label{eq:deformation_fac}
\prod_{k=2,4,6} \left(1 + b_k\beta_k^2\right) \, ,
\end{align}
consisting of deformation correction coefficients \cite{Wang2010b}, 
\begin{align}
\label{eq:deformation_coeff}
b_k = \frac{k}{2} g_1 A^{1/3} + \left(\frac{k}{2}\right)^2 g_2 A^{-1/3} \, ,
\end{align}
with $g_1$ and $g_2$ coupled with the first and second terms, respectively. 
The macroscopic part, \(B_i^{\text{macro}}\), consists of the volume, surface, Coulomb, symmetry, pairing, exchange-Coulomb, and curvature terms of the BWN model, 
\begin{align}
\label{eq:BWL_macro}
B_i^{\text{macro}} = &~\alpha_\mathrm{v} A + \alpha_\mathrm{s} A^{2/3} + \alpha_\mathrm{c} Z^2 A^{-1/3} + \alpha_{\mathrm{sym},I} I^2 A f_\mathrm{s} \nonumber \\
&~+ \delta_\mathrm{np} \alpha_\mathrm{p} A^{-1/3} + \alpha_\mathrm{xc} Z^{4/3} A^{-1/3} + \alpha_\mathrm{r} A^{1/3} \, , 
\end{align}
whereas the phenomenological shell correction term, \(B_i^{\text{shell}}\), regulated by the deformation quenching factor, 
\begin{equation}
\label{eq:deformation_quench}
\exp\left(-\gamma \sum_k \beta_k^2\right) \,  ,
\end{equation}
reads
\begin{equation}
\label{eq:BWL_shell}
B_i^{\text{shell}} = \alpha_\mathrm{m} P + \beta_\mathrm{m} P^2 + c_\mathrm{m} (v_\mathrm{n} + v_\mathrm{p}) + e_\mathrm{m1} \delta_{\mathrm{shell}} \, e^{e_\mathrm{m2} (v_\mathrm{p}^2 + v_\mathrm{n}^2)} \, .
\end{equation}
Hereinafter, the BWN mass model embedded with the above deformation and phenomenological shell corrections is referred to as the BWL model.

\subsection{Bayesian inference with\\Markov chain Monte Carlo sampling approach}
\label{sec:BA_MCMC}

We construct a Bayesian analysis (BA) framework with the Markov chain Monte Carlo (MCMC) sampling approach to study the correlations of mass-model parameters of the BW variants, i.e., BW, BWK, BWN, and BWL. We first take the BWL model as a case study and then generalize the BA-MCMC framework for the BW, BWK, and BWN mass models.

\subsubsection{Bayes' theorem, model parameters and data set}

The posterior probability distribution of the parameters, $\boldsymbol{\theta}$, of the mass model with the selected experimental data set, $\mathcal{D}$, follows Bayes' theorem (e.g., Ref.~\cite{MacKay2003}), 
\begin{equation}
\label{eq:Bayes}
p(\boldsymbol{\theta} \mid \mathcal{D}) = \frac{p(\mathcal{D} \mid \boldsymbol{\theta}) \, p(\boldsymbol{\theta})}{p(\mathcal{D})} \, ,
\end{equation}
where $\boldsymbol{\theta} = (\alpha_\mathrm{v}, \alpha_\mathrm{s}, \alpha_\mathrm{c}, \ldots, g_1, g_2, \gamma)$ is the 18-dimensional parameter vector of the BWL mass model, c.f., Eqs.~(\ref{eq:BWL}), (\ref{eq:BWL_macro}), and (\ref{eq:BWL_shell}). For the BW, BWK, and BWN models, the respective $\boldsymbol{\theta}$ spaces are listed in Table~\ref{tab:prior_initial}. 

In conventional optimization or bounded Bayesian analysis, the noise parameter, $\sigma$, is often fixed to a constant value. To avoid the potential underestimation of theoretical uncertainties, we develop a full Bayesian framework that jointly infers both the mass-model parameters, $\boldsymbol{\theta}$, and the noise standard deviation, $\sigma$. Consequently, the parameter vector for the BWL model is extended to a 19-dimensional space, expressed as
\begin{equation}
\boldsymbol{\theta}_{\text{full}} = (\alpha_\mathrm{v}, \alpha_\mathrm{s}, \alpha_\mathrm{c}, \dots, \gamma, \log\sigma) \, ,
\end{equation}
where $\log\sigma$ denotes the logarithm of the noise standard deviation. Treating $\sigma$ as an unknown free parameter allows the uncertainty in the intrinsic noise level of the experimental data to be fully propagated into the final posterior distribution. In practice, we sample $\log\sigma$ to naturally enforce the physical constraint $\sigma > 0$.

The selected experimental dataset is defined as $\mathcal{D} = \{B_i^{\mathrm{Exp}}(x_i)\}_{i=1}^n$, where $B_i^{\mathrm{Exp}}$ denotes the experimental binding energy for nucleus $i$, and $x_i = (Z_i, A_i, \beta_{2,i}, \beta_{4,i}, \beta_{6,i})$ comprises the respective proton number, mass number, and theoretical deformation parameters $\beta_2$, $\beta_4$, and $\beta_6$ used in the WS4 mass model~\cite{WS4database}. In total, we select $n=2242$ experimental binding energies with uncertainty lower than 100~keV from the intersection of the AME2020 and WS4 compilations~\cite{AME2020a,WS4database} to form $\mathcal{D}$. In Eq.~(\ref{eq:Bayes}), $p(\mathcal{D} \mid \boldsymbol{\theta}_{\text{full}})$ represents the likelihood function, $p(\boldsymbol{\theta}_{\text{full}})$ is the prior probability distribution, and
\begin{equation}
p(\mathcal{D}) = \int p(\mathcal{D}\mid \boldsymbol{\theta}_{\text{full}}) p(\boldsymbol{\theta}_{\text{full}}) \, d\boldsymbol{\theta}_{\text{full}}
\end{equation}
denotes the marginal likelihood (evidence).

\begin{table}[t!]
\caption{\label{tab:prior_initial}%
Prior initial values (parameters) of BW, BWK, BWN, and BWL.}
\begin{tabular*}{\linewidth}{@{\hspace{5mm}\extracolsep{\fill}}crrrr@{\hspace{5mm}}}
\toprule[1.0pt]
\midrule[0.25pt]
Parameter & BW\footnotemark[1] & BWK\footnotemark[1] & BWN\footnotemark[1] & BWL\footnotemark[2]\\
\midrule[0.25pt]
$\alpha_\mathrm{v}$   & $15.5255 $ & $16.4920 $ & $16.7043 $ & $16.7400 $ \\
$\alpha_\mathrm{s}$   & $-16.8949$ & $-25.5618$ & $-26.3000$ & $-26.3000$ \\
$\alpha_\mathrm{c}$   & $-0.7022 $ & $-0.7614 $ & $-0.7615 $ & $-0.7600$ \\
$\alpha_\mathrm{sym}$ & $-22.9874$ & $-32.5777$ & --         & -- \\
$c_\mathrm{sym}$      & --         & --         & $-35.3636$ & $-35.9000$ \\
$\alpha_\mathrm{xc}$  & --         & $1.6997  $ & $5.9751  $ & $5.9300  $ \\
$\alpha_\mathrm{w}$   & --         & $-61.7229$ & --         & -- \\
$\alpha_\mathrm{p}$   & --         & $11.0409 $ & $1.4405  $ & $1.3900  $ \\
$\alpha_\mathrm{r}$   & --         & $13.3315 $ & $14.1287 $ & $14.2000 $ \\
$\alpha_\mathrm{st}$  & --         & $61.1172 $ & --         & -- \\
$\alpha_\mathrm{m}$   & --         & $-2.0293 $ & $-1.0877 $ & $-0.7700 $ \\
$\beta_\mathrm{m}$    & --         & $-2.0293  $ & $0.1615  $ & $0.1120  $ \\
$c_\mathrm{m}$        & --         & --         & $-0.2343 $ & $-0.2650 $ \\
$e_\mathrm{m1}$       & --         & --         & $5.4713  $ & $5.2600  $ \\
$e_\mathrm{m2}$       & --         & --         & $-0.0444 $ & $-0.0457 $ \\
$k$                   & --         & --         & $2.0829  $ & $2.1400  $ \\
$\xi$                 & --         & --         & $1.2216  $ & $1.2900  $ \\
$\kappa_\mathrm{s}$   & --         & --         & $0.2491  $ & $0.1800  $ \\
$g_\mathrm{1}$        & --         & --         & --         & $0.0200  $ \\
$g_\mathrm{2}$        & --         & --         & --         & $-0.700  $ \\
$\gamma$              & --         & --         & --         & $3.5500  $ \\
$\log\sigma$\footnotemark[3] & 0.0 & 0.0        & 0.0        & $0.7000  $ \\
\bottomrule[1.0pt]
\end{tabular*}
\footnotetext[1]{Prior initial values obtained by \citet{Wu2025} using the least-squares fit method.}
\footnotetext[2]{Prior initial values deduced in this work using Bayesian analysis with MCMC sampling.}
\footnotetext[3]{The initial prior values of $\log\sigma$ specified in this work.}
\end{table}

\begin{table}[t!]
\caption{\label{tab:prior_std_dev}%
Prior standard deviations of all parameters of BW, BWK, BWN, and BWL.}
\begin{tabular*}{\linewidth}{@{\hspace{5mm}\extracolsep{\fill}}ccccc@{\hspace{5mm}}}
\toprule[1.0pt]
\midrule[0.25pt]
Parameter & BW & BWK & BWN & BWL\\
\midrule[0.25pt]
$\alpha_\mathrm{v}$   & 0.2500 & 0.2500 & 0.2500 & 0.2500 \\
$\alpha_\mathrm{s}$   & 0.7500 & 0.5000 & 0.7500 & 0.7500 \\
$\alpha_\mathrm{c}$   & 0.1000 & 0.1000 & 0.1000 & 0.1000 \\
$\alpha_\mathrm{sym}$ & 0.7500 & 0.5000 & --     & --     \\
$c_\mathrm{sym}$      & --     & --     & 1.0000 & 1.0000 \\
$\alpha_\mathrm{xc}$  & --     & 0.1500 & 0.2500 & 0.2500 \\
$\alpha_\mathrm{w}$   & --     & 3.7500 & --     & --     \\
$\alpha_\mathrm{p}$   & --     & 0.7500 & 0.2500 & 0.5000 \\
$\alpha_\mathrm{r}$   & --     & 0.8750 & 0.7500 & 1.0000 \\
$\alpha_\mathrm{st}$  & --     & 3.7500 & --     & --     \\
$\alpha_\mathrm{m}$   & --     & 0.2500 & 0.2500 & 0.1250 \\
$\beta_\mathrm{m}$    & --     & 0.1000 & 0.1000 & 0.1000 \\
$c_\mathrm{m}$        & --     & --     & 0.1000 & 0.1000 \\
$e_\mathrm{m1}$       & --     & --     & 0.2500 & 1.2500 \\
$e_\mathrm{m2}$       & --     & --     & 0.1000 & 0.1000 \\
$k$                   & --     & --     & 0.2500 & 0.5000 \\
$\xi$                 & --     & --     & 0.2500 & 0.2500 \\
$\kappa_\mathrm{s}$   & --     & --     & 0.1000 & 0.1000 \\
$g_\mathrm{1}$        & --     & --     & --     & 0.1000 \\
$g_\mathrm{2}$        & --     & --     & --     & 0.3500 \\
$\gamma$              & --     & --     & --     & 1.0000 \\
$\log\sigma$\footnotemark[1]  & 0.5500 & 0.3750 & 0.2500 & 0.2750 \\
\bottomrule[1.0pt]
\end{tabular*}
\footnotetext[1]{The prior standard deviation assigned to the noise parameter $\log\sigma$ in this work.}
\end{table}

\subsubsection{Likelihood function and full Bayesian treatment}

The likelihood function assumes that the model residuals (the differences or deviations between experimental and theoretical binding energies) are independent and identically distributed according to Gaussian uncertainty, 
\begin{equation}
p(\mathcal{D} \mid \boldsymbol{\theta}_{\text{full}}) = \prod_{i=1}^{n} \mathcal{N}\left(B_i^{\mathrm{Exp}} \mid B_i^{\mathrm{Th}}(\boldsymbol{\theta}_{\text{full}}), \sigma^2\right) \, ,
\end{equation}
where $\sigma$ is implicitly contained within $\boldsymbol{\theta}_{\text{full}}$. In logarithmic form, the log-likelihood reads
\begin{align}
\log p(\mathcal{D} \mid \boldsymbol{\theta}_{\text{full}}) = & -\frac{n}{2} \log(2\pi\sigma^2) \nonumber \\
& - \frac{1}{2\sigma^2} \sum_{i=1}^{n} \left[B_i^{\mathrm{Exp}} - B_i^{\mathrm{Th}}(\boldsymbol{\theta}_{\text{full}})\right]^2 \, .
\end{align}

\subsubsection{Prior distribution}

The prior probability distribution $p(\boldsymbol{\theta}_{\text{full}})$ for the parameters of the multi-variant mass models incorporates physically motivated boundary constraints derived from previous works~\cite{Kirson2008, Wu2025} and nuclear mass systematics. To establish reliable bounds within the full Bayesian framework, initial single-chain MCMC simulations under conservative uniform priors are performed to thoroughly scan the parameter space and exclude non-physical regions. Based on these exploratory runs, the joint prior distribution is constructed by assigning an independent Gaussian distribution to each parameter, including the noise parameter $\log\sigma$, 
\begin{equation}
p(\boldsymbol{\theta}_{\text{full}}) = \prod_{j=1}^{d} \mathcal{N}(\theta_j \mid \mu_j, \sigma_j^2) \, ,
\end{equation}
where $\mu_j$ represents the prior mean, $\sigma_j$ is the prior standard deviation, and $d$ denotes the total number of inferred parameters ($d=19$ for the BWL model, and varies for the other BW variants according to Table~\ref{tab:prior_initial}). The prior means, $\mu_j$, are self-consistently determined through the multi-chain MCMC framework discussed in Sec.~\ref{sec:MultiChain} later.

The prior initial values ($\theta_j^{\text{initial}}$) and prior standard deviations ($\sigma_j^{\text{prior}}$) for the BW, BWK, BWN, and BWL mass models are summarized in Table~\ref{tab:prior_initial} and Table~\ref{tab:prior_std_dev}, respectively. To ensure stable exploration and prevent frequent rejections near boundary constraints during the initial sampling stage, the base step size ($\boldsymbol{\sigma}_{\mathrm{base}}$) of the random-walk proposal distribution is defined as $0.005 \times \sigma_j^{\mathrm{prior}}$.

\subsubsection{Posterior distribution}

Combining the likelihood function and the joint prior distribution implemented in the BA-MCMC framework, the unnormalized log-posterior distribution for the extended 19-dimensional parameter space $\boldsymbol{\theta}_{\text{full}}$ for BWL is formulated as 
\begin{align}
\log p(\boldsymbol{\theta}_{\mathrm{full}} \mid \mathcal{D}) \propto &~ \log p(\mathcal{D} \mid \boldsymbol{\theta}_{\mathrm{full}}) + \log p(\boldsymbol{\theta}_{\mathrm{full}}) \nonumber \\
= &~ -\frac{n}{2} \log(2\pi\sigma^2) \nonumber \\
&~ - \frac{1}{2\sigma^2} \sum_{i=1}^{n} \left[B_i^{\mathrm{Exp}} - B_i^{\mathrm{Th}}(\boldsymbol{\theta}_{\mathrm{full}})\right]^2 \nonumber \\
&~- \frac{1}{2} \sum_{j=1}^{d} \left( \frac{\theta_j - \mu_j}{\sigma_j^{\mathrm{prior}}} \right)^2  \nonumber \\
&~ + \sum_{k \in \mathcal{C}} \log \mathbb{I}\left[\theta_k \in \mathcal{M}_{\mathrm{phys}}\right] \, ,
\end{align}
where $d$ represents the total number of inferred parameters (including $\log\sigma$). The third term on the right-hand side represents the joint log-Gaussian prior density kernel. The fourth term utilizes the indicator function $\mathbb{I}[\cdot]$ to enforce the physically motivated boundary constraints $\mathcal{C}$ (e.g., ensuring the predefined macro-model coefficients to remain negative and preventing non-positive exponents), $\mathcal{M}_{\text{phys}}$ denotes the physically allowable domain enforced by the joint boundary constraints $\mathcal{C}$, which returns $1$ if all constraints are satisfied and $0$ (yielding $-\infty$ in the logarithm) otherwise, matching the rejection logic in the numerical pipeline.

\subsubsection{Markov chain Monte Carlo sampling}
\label{sec:MCMC}

For posterior sampling, we implement the Metropolis-Hastings algorithm \cite{metropolis1953, hastings1970, chib1995}. The acceptance probability for a proposed parameter vector $\boldsymbol{\theta}^*$ is 
\begin{equation}
\alpha(\boldsymbol{\theta}^* \mid \boldsymbol{\theta}^{(t)}) = \min\left\{1, \frac{p(\boldsymbol{\theta}^* \mid \mathcal{D})}{p(\boldsymbol{\theta}^{(t)} \mid \mathcal{D})}\right\} \, .
\end{equation}
An adaptive Gaussian proposal distribution is used 
\begin{equation}
q(\boldsymbol{\theta}^* \mid \boldsymbol{\theta}^{(t)}) = \mathcal{N}(\boldsymbol{\theta}^* \mid \boldsymbol{\theta}^{(t)}, \boldsymbol{\Sigma}) \, ,
\end{equation}
where $t$ denotes the last iteration index, $\mathcal{N}$ is the probability density function, covariance matrix $\boldsymbol{\Sigma}$ is adaptively tuned during the burn-in phase to achieve the optimal acceptance rate of $0.234$ for high-dimensional sampling, referring to the optimization scaling factor of the random walk of Metropolis algorithms proposed by \citet{Roberts1997}. They implemented weak convergence theory to demonstrate that, for high-dimensional target distributions with independent and identically distributed components, the efficiency of the algorithm is maximized when the acceptance rate approaches approximately 0.234. Subsequent studies have shown that this result is remarkably robust holding under much more general conditions far beyond the original independent and identically distributed product form, and can be extended to various proposal distributions and even multimodal target distributions. In practice, their theoretical findings provide a concrete guideline for tuning the proposed step size of the MCMC random walk.

During sampling, we introduce a moving‑window of average acceptance rate, of which the acceptance outcome of a fixed number of recent iterations is recorded, the mean acceptance rate in the window is periodically computed, and a global scaling factor $s$ is updated by exponential smoothing. This dynamically adjusts the scale of the proposed step size. To further stabilize the adaptation, we decompose the proposal standard deviation into a base step size $\boldsymbol{\sigma}_{\text{base}}$ and the scaling factor $s$ as $\boldsymbol{\sigma}_{\text{prop}} = s \cdot \boldsymbol{\sigma}_{\text{base}}$.

We initially set the base step size to a small multiple of the prior standard deviation with $\boldsymbol{\sigma}_{\text{base}} = 0.005 \times \sigma_j^{\text{prior}}$, to avoid frequent rejections caused by prior boundaries. Meanwhile, we employ an initial warm‑up period to sample the chain with a fixed step size without adaptation, ensuring that the chain first explores and migrates into the high‑probability region of the posterior before adaptation begins. Subsequently, during the adaptive burn‑in phase, we periodically update $\boldsymbol{\sigma}_{\text{base}}$ using the covariance matrix estimated from the most recent samples. Following the theoretical approach of Refs.~\cite{Roberts1997, Gelman1996}, the empirical covariance is scaled by a factor of $2.38/\sqrt{d}$, which maximizes the efficiency of mixing when the target acceptance rate is close to 0.234. This choice has been shown to be robust under more general conditions~\cite{Schmon2021, li2025}.

The adaptive random-walk Metropolis algorithm is used with the following configurations, (i) the target acceptance rate is 0.234, (ii) the moving window size for acceptance monitoring is 200, (iii) learning rate for scaling factor update is 0.02, (iv) the scaling factor $ s = 0.02$-2.0, (v) the number of warm-up steps is 100,000, (vi) the base step size factor is initialized as 0.005 to scale the prior standard deviations. For the covariance adaptation interval, in every 20,000 steps, the covariance estimation is based on the most recent 2000 samples with the theoretical scaling factor of $ 2.38/\sqrt{d} $.

\subsubsection{Multi-chain MCMC framework}
\label{sec:MultiChain}

To ensure exhaustive exploration across the entire parameter landscape and to obtain high reliability of the statistical optimization, we deploy a parallel multi-chain MCMC framework with standard unbiased configurations. We initialize $n_c = 4$ independent Markov chains, where the initial state of the $m$-th chain ($\boldsymbol{\theta}_{m}^{(0)}$) is sampled globally across the parameter space, 
\begin{equation}
\boldsymbol{\theta}_{m}^{(0)} \sim \text{Prior}(\boldsymbol{\theta}) \, ,
\end{equation}
where $m = 1, 2, \dots, n_c$ denotes the chain index. To ensure distinct and independent stochastic trajectories, each chain is assigned a separate random seed incremented by a step (${\text{Seed}}_m = 42 + m \times 1000$). Within this parallel framework, the first chain ($m = 1$) is designated as the reference chain, which is maintained at the target distribution and is utilized for the final posterior inference. The remaining chains act as auxiliary chains to facilitate a cross-chain convergence diagnosis. During the sampling process, the proposal distribution is regularly updated based on the covariance matrix of the collected samples to further improve efficiency.

We use the Gelman-Rubin statistics, $\hat{R}$, to verify the convergence of sampling by comparing the variation between different independent Markov chains with the variation within each chain, 
\begin{equation}
\label{eq:GelmanRubin}
\hat{R} = \sqrt{\frac{\hat{V}}{W}} \, ,
\end{equation}
where $\hat{V}$ denotes the overall variance (pool between-chain and within-chain variance), and $W$ represents the average within-chain variance \cite{Gelman1992, Gelman2021}. The sampling achieves convergence when $\hat{R} < 1.1$ \cite{Gelman2021}.

Meanwhile, we also verify the effective sample size ($n_\mathrm{eff}$) to gauge the sampling efficiency by estimating the number of effectively independent samples \cite{Robert1998, 2019Biome.106..321V}, 
\begin{equation}
\label{eq:neff}
n_\mathrm{eff} = \frac{N_{\text{total}}}{1 + 2 \sum_{j=1}^{M} \rho_j} \, ,
\end{equation}
where $N_{\text{total}}$ is the total number of samples analyzed, $\rho_j$ is the autocorrelation coefficient at lag $j$, and $M$ is the cutoff lag beyond which autocorrelations become negligible \cite{Ripley1987}. An $n_\mathrm{eff}>400$ is considered adequate for reliable inference \cite{Gelman2021}.

We perform posterior sampling for the BW, BWK, BWN, and BWL mass models using multi-chain parallel MCMC. We run four parallel chains, each with $ 10^7 $ iterations. We discard the first $ 5\times10^6 $ iterations as burn-in to eliminate the influence of initial values. We then use the second $ 5\times10^6 $ samples from the reference chain for posterior analysis.

\subsubsection{Parameter Correlation and Degeneracy Diagnosis}
\label{sec:CorrelationDiagnosis}

We employ the Spearman rank correlation coefficient~\cite{gelman2013bayesian} to diagnose and quantify parameter correlations and degeneracies of the multi-dimensional parameter spaces evaluated from the MCMC posterior samples for nuclear mass models. The Spearman rank correlation coefficient, 
\begin{equation}
\label{eq:Spearman}
\rho_{ij} = 1 - \frac{6 \sum_{k=1}^{n_{\mathrm{diag}}} d_k^2}{n_{\mathrm{diag}}(n_{\mathrm{diag}}^2 - 1)},
\end{equation}
where $d_k = \text{R}[\theta_{i}]_k - \text{R}[\theta_{j}]_k$ represents the difference between the $\text{R}[\theta_{i}]$ and $\text{R}[\theta_{j}]$ ranks of the $k$-th MCMC sample for the two parameters, and $n_{\mathrm{diag}}$ is the total number of samples utilized for diagnosis. Parameter degeneracy within the entire fitting matrix is monitored by the condition number of a given Spearman rank correlation matrix $\mathbf{R}_\mathrm{S}$, 
\begin{equation}
\label{eq:condition_number}
\mathcal{K}(\mathbf{R}_\mathrm{S}) = \frac{\lambda_{\max}}{\lambda_{\min}} \, ,
\end{equation}
where $\lambda_{\max}$ and $\lambda_{\min}$ are the largest and smallest eigenvalues of $\mathbf{R}_\mathrm{S}$, respectively. Typically, a condition number $\mathcal{K}(\mathbf{R}_\mathrm{S}) > 10^3$ indicates a high degree of parameter degeneracy. Meanwhile, we also calculate the Pearson correlation coefficient, $r_{ij}$, between parameters $\theta_i$ and $\theta_j$, quantifying the strength of the linear correlation,
\begin{equation}
\label{eq:Pearson}
r_{ij} = \frac{\text{cov}(\theta_i, \theta_j)}{\sigma_{\theta_i} \sigma_{\theta_j}},
\end{equation}
where $\text{cov}(\theta_i, \theta_j)$ denotes the covariance, and $\sigma_{\theta_i}$ ($\sigma_{\theta_j}$) represents the standard deviation of the $\theta_i$ ($\theta_j$) posterior distribution. The formulation of the condition number of a given Pearson correlation matrix, $\mathcal{K}(\mathbf{R}_\mathrm{P})$, is similar to Eq.~(\ref{eq:condition_number}).

As the nuclear mass model becomes increasingly sophisticated (e.g., incorporating highly nonlinear shell corrections or deformation quenching effects), the resulting joint posterior distributions could deviate from ideal multivariate Gaussian profiles, exhibiting complex nonlinear manifolds and heavy tails. A notable discrepancy between $\rho_{ij}$ and $r_{ij}$ serves as a diagnostic signal indicating a prominent departure from linearity in the parameter manifold.

\subsubsection{Posterior predictive distribution}
\label{sec:PPD}

The posterior predictive distribution for a new set of binding energies $B^{\mathrm{new}}$ given by the observed data $\mathcal{D}$ is defined as 
\begin{equation}
\label{eq:posterior_predict_dist}
p(B^{\mathrm{new}} \mid \mathcal{D}) = \int p(B^{\mathrm{new}} \mid \boldsymbol{\theta}_{\text{full}}) \, p(\boldsymbol{\theta}_{\text{full}} \mid \mathcal{D}) \, d\boldsymbol{\theta}_{\text{full}} \, .
\end{equation}
In practice, the integral of Eq.~(\ref{eq:posterior_predict_dist}) can be numerically approximated using the collected Monte Carlo samples, 
\begin{equation}
\mathbb{E}[B^{\mathrm{new}} \mid \mathcal{D}] \approx \frac{1}{S} \sum_{s=1}^{S} B^{\mathrm{Th}}(\boldsymbol{\theta}^{(s)}) \, ,
\end{equation} 
where $\{\boldsymbol{\theta}^{(s)}\}_{s=1}^{S}$ are the valid posterior parameter samples generated from the multi-chain MCMC sampler (specifically extracted from the reference chain, see Sec.~\ref{sec:MultiChain} for the details), and $S$ denotes the total number of samples used for inference. Eventually, the overall uncertainty associated with the posterior predictive distribution, reflecting the discrepancy between the theoretical mass model predictions and experimental data, is quantified by integrating both the parameter uncertainty and the intrinsic data noise, 
\begin{align}
\sigma^2(B^{\mathrm{new}}) = &~ \frac{1}{S} \sum_{s=1}^{S} \left[ B^{\mathrm{Th}}(\boldsymbol{\theta}^{(s)}) - \mathbb{E}[B^{\mathrm{new}} \mid \mathcal{D}] \right]^2 \nonumber \\
&~ + \frac{1}{S} \sum_{s=1}^{S} \left(\sigma^{(s)}\right)^2 \, ,
\end{align}
where $\sigma^{(s)} = \exp(\theta_{\text{log}\sigma}^{(s)})$ represents the noise standard deviation inferred at the $s$-th posterior sample.

For the subsequent discussions and graphical representations, the total theoretical uncertainty for a given nucleus is quantified by the posterior predictive standard deviation, $\sigma(B)$, which is obtained via the square root transformation, 
\begin{equation}
\label{ref:sigma(b)}
\sigma(B) = \sqrt{\sigma^2(B^{\mathrm{new}})} \, .
\end{equation}

\section{Results and Discussions of BA-MCMC Analysis}

All model parameters, prior initial values, and statistical bounds in this work are consistently reported up to three or four decimal places. This precision aligns with most of the experimental resolution of nuclear binding energies in the AME2020 compilation. Adherence to this convention prevents the introduction of physically unjustified spurious precision in high-dimensional statistical inference. The new parameter sets of BW, BWK, BWN, and BWL deduced in this work using BA-MCMC are labeled with an asterisk (*), while the optimal parameter sets obtained from the lowest RMS over all MCMC samples are labeled with a dagger ($\dagger$) symbol.

\begin{figure*}
\hspace{-27em}
\includegraphics[width=0.5\linewidth]{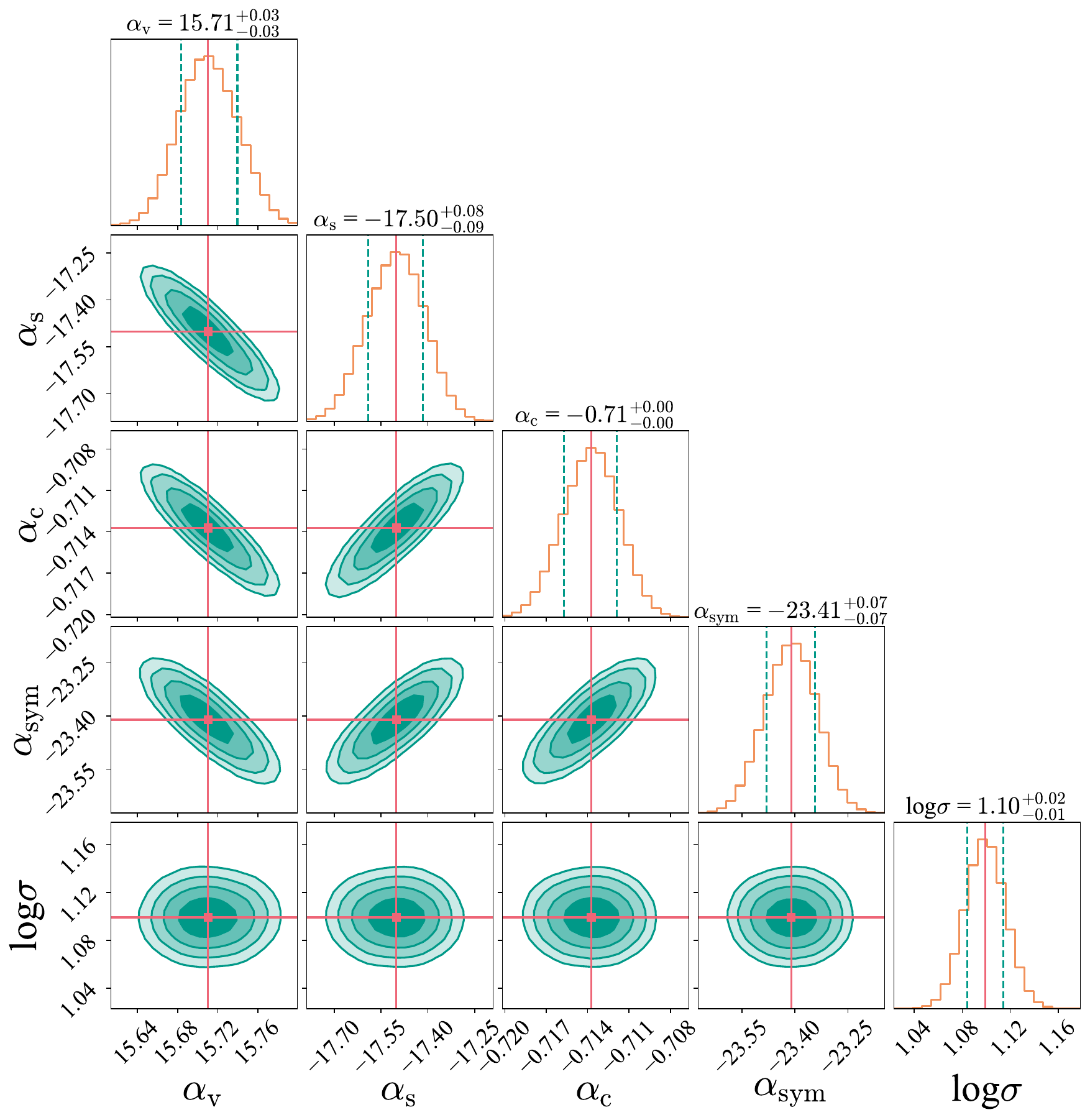}%
\makebox[0pt][r]{%
\raisebox{12em}{%
\includegraphics[width=0.6\linewidth]{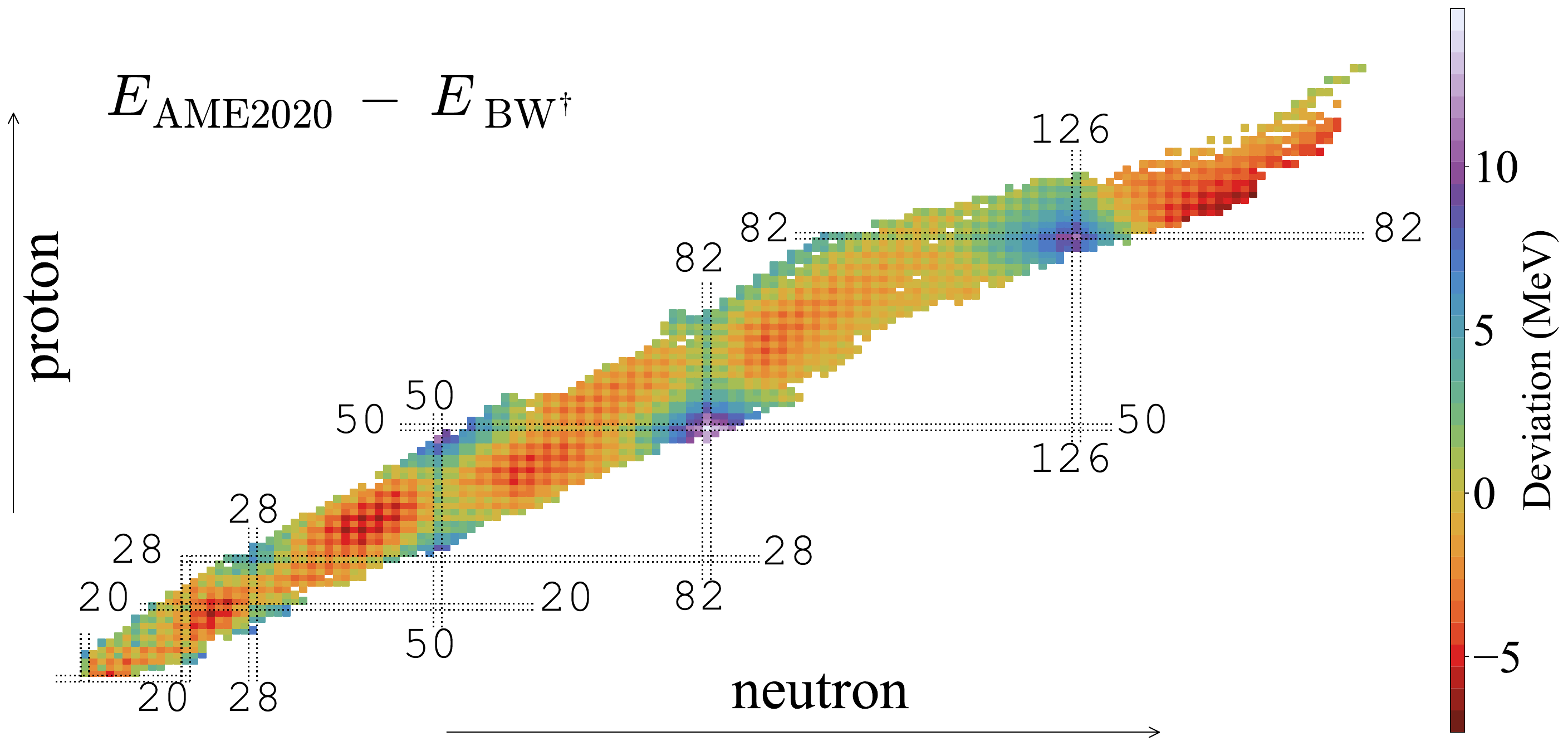}%
}\hspace*{-28em}%
}%
\caption{\label{fig:posterior_BW}Posterior distributions for 4 parameters of the BW* nuclear mass model. The 2-dimensional (2D) contour levels indicate the 38\%, 68\%, 87\%, and 95\% credible regions. The one-dimensional (1D) marginalized posteriors are diagonally plotted with the 68\% credible interval shade. The maximum likelihood estimates for the 1D posteriors of each parameter are summarized above the respective subplot. The most probable distribution of each parameter of BW* is shown on top of the respective diagonal plot. The mean values of posterior distributions of all 4 parameters along with the corresponding physical units of BW* are listed in the third column of Table~\ref{tab:mass_models_01}. The optimal parameter set along with the corresponding physical units is presented in the fourth column of Table~\ref{tab:mass_models_01}. The deviation of BW$^\dagger$ and AME2020 binding energies is presented at the nuclear chart with respect to the right color scale.}
\end{figure*}

\subsection{BW mass model}
\label{sec:BW_MCMC}

Table~\ref{tab:mass_model_BW} summarizes the effective sample size, $n_\mathrm{eff}$, and the Gelman-Rubin statistics, $\hat{R}$, for all parameters of the BW* mass model.  All $\hat{R}$ values are below 1.0001, indicating convergence, and the $n_\mathrm{eff}$ values exceed 2600, ensuring reliable posterior estimates. The posterior mean of the noise standard deviation $\sigma$ is 3.025~MeV with a 95~\% credible interval of [2.94, 3.11]~MeV.

\begin{table}[t!]
\caption{\label{tab:mass_model_BW}%
Convergence diagnosis for the reference chain of BW*. See Sec.~\ref{sec:MultiChain}, Eqs.~(\ref{eq:GelmanRubin}) and (\ref{eq:neff}) for the details.}
\begin{tabular*}{\linewidth}{@{\hspace{10mm}\extracolsep{\fill}}ccc@{\hspace{10mm}}}
\toprule[1.0pt]
\midrule[0.25pt]Parameter & $n_\mathrm{eff}$ & $\hat{R}$\\
\midrule[0.25pt]
$\alpha_\mathrm{v}$   & 2679.2 & 1.0000 \\
$\alpha_\mathrm{s}$   & 2716.8 & 1.0000 \\
$\alpha_\mathrm{c}$   & 2743.3 & 1.0000 \\
$\alpha_\mathrm{sym}$ & 2976.3 & 1.0000 \\
$\log\sigma$          & 9205.6 & 1.0000 \\
\bottomrule[1.0pt]
\end{tabular*}
\end{table}

\begin{table}[t!]
\caption{\label{tab:bw_correlation}%
Pearson correlation matrix and Spearman rank correlation matrix of the posterior MCMC samples for the BW* nuclear mass model parameters. See Sec.~\ref{sec:CorrelationDiagnosis}, Eqs.~(\ref{eq:Spearman}) and (\ref{eq:Pearson}) for the details.}
\begin{tabular*}{\linewidth}{@{\hspace{5mm}\extracolsep{\fill}}crrrrr@{\hspace{5mm}}}
\toprule[1.0pt]
\midrule[0.25pt]
\multicolumn{6}{c}{Pearson correlation matrix}\\
 & $\alpha_\mathrm{v}$ & $\alpha_\mathrm{s}$ & $\alpha_\mathrm{c}$ & $\alpha_{\rm sym}$ & $\log\sigma$ \\
\midrule[0.25pt]
$\alpha_\mathrm{v}$     & 1.000 & $-0.994$ & $-0.986$ & $-0.926$ & 0.005 \\
$\alpha_\mathrm{s}$     & $-0.994$ & 1.000 & 0.966 & 0.912 & $-0.005$ \\
$\alpha_\mathrm{c}$     & $-0.986$ & 0.966 & 1.000 & 0.895 & $-0.005$ \\
$\alpha_{\rm sym}$ & $-0.926$ & 0.912 & 0.895 & 1.000 & $-0.002$ \\
$\log\sigma$   & 0.005 & $-0.005$ & $-0.005$ & $-0.002$ & 1.000 \\

\toprule[1.0pt]
\midrule[0.25pt]
\multicolumn{6}{c}{Spearman rank correlation matrix}\\
 & $\alpha_\mathrm{v}$ & $\alpha_\mathrm{s}$ & $\alpha_\mathrm{c}$ & $\alpha_{\rm sym}$ & $\log\sigma$ \\
\midrule[0.25pt]
$\alpha_\mathrm{v}$     & 1.000 & $-0.994$ & $-0.986$ & $-0.923$ & 0.006 \\
$\alpha_\mathrm{s}$     & $-0.994$ & 1.000 & 0.964 & 0.909 & $-0.006$ \\
$\alpha_\mathrm{c}$     & $-0.986$ & 0.964 & 1.000 & 0.891 & $-0.006$ \\
$\alpha_{\rm sym}$ & $-0.923$ & 0.909 & 0.891 & 1.000 & $-0.004$ \\
$\log\sigma$   & 0.006 & $-0.006$ & $-0.006$ & $-0.004$ & 1.000 \\
\bottomrule[1.0pt]
\end{tabular*}
\end{table}%

With the large data set of 2242 nuclei providing strong statistical constraints, the marginal posteriors of the BW* model is naturally produced in a highly symmetrical shape (diagonal plots of Gaussian shapes in Fig.~\ref{fig:posterior_BW}). The posterior distributions for all parameters of the BW* model exhibit a symmetric and unimodal Gaussian shape, with the posterior means concentrated within narrow ranges. The symmetrical parameter space is mainly due to (i) the four parameters ($  \alpha_\mathrm{v}  $, $  \alpha_\mathrm{s}  $, $  \alpha_\mathrm{c}  $, $  \alpha_{\rm sym}  $) of the BW* mass model are linearly correlated \cite{Lundquist2020, Saito2024}, rendering the likelihood function equivalent to a multivariate linear regression, (ii) the adoption of Gaussian priors combined with a Gaussian likelihood causes the joint posterior mathematically a multivariate Gaussian distribution \cite{Niu2018}, and (3) the absence of strong nonlinear terms or stringent physical constraints. The two-dimensional contour plots in Fig.~\ref{fig:posterior_BW} further reveal an inverse correlation between $ \alpha_\mathrm{v} $-$ \alpha_\mathrm{s} $, $ \alpha_\mathrm{v} $-$ \alpha_\mathrm{c} $, and $ \alpha_\mathrm{v} $-$ \alpha_{\rm sym} $, while $  \alpha_\mathrm{s}  $, $  \alpha_\mathrm{c}  $, and $  \alpha_{\rm sym}  $ are positively correlated with each other.

The detailed correlation strengths between the model parameters are presented in the Pearson correlation and Spearman rank correlation matrices in Table~\ref{tab:bw_correlation}. We take an example of the Spearman correlation coefficient of $\alpha_\mathrm{v}$-$\alpha_\mathrm{s}$, corresponding to $\rho_{ij}$ in Eq.~(\ref{eq:Spearman}) with $\theta_i=\alpha_\mathrm{v}$ and $\theta_j=\alpha_\mathrm{s}$, i.e., $\rho(\alpha_\mathrm{v},\alpha_\mathrm{s}) = -0.994$. The Pearson coefficients are presented analogously, with $r(\alpha_\mathrm{v},\alpha_\mathrm{s})$ denoting $r_{ij}$ for the same pair of posterior parameters in Eq.~(\ref{eq:Pearson}). Both matrices of the Spearman correlation coefficient and Pearson correlation coefficient are symmetric. The Spearman coefficients of $\rho_{ij} > 0$ and $\rho_{ij} < 0$ indicate positive and negative correlations among the posterior parameters, respectively. These correlation patterns are consistent with the physical expectations of the model (Fig.~\ref{fig:posterior_BW}). Furthermore, the joint distributions of $  \sigma  $ with each parameter appear approximately circular, suggesting uniform correlations.

\begin{table}[t!]
\caption{\label{tab:mass_model_BWK}%
Convergence diagnosis for the reference chain of BWK*. See Sec.~\ref{sec:MultiChain}, Eqs.~(\ref{eq:GelmanRubin}) and (\ref{eq:neff}) for the details.}
\begin{tabular*}{\linewidth}{@{\hspace{10mm}\extracolsep{\fill}}ccc@{\hspace{10mm}}}
\toprule[1.0pt]
\midrule[0.25pt]
Parameter & $n_\mathrm{eff}$  & $\hat{R}$ \\
\midrule[0.25pt]
$\alpha_\mathrm{v}$    & 2527.9 & 1.0008 \\
$\alpha_\mathrm{s}$    & 2520.1 & 1.0010 \\
$\alpha_\mathrm{c}$    & 2571.5 & 1.0001 \\
$\alpha_\mathrm{sym}$  & 2559.6 & 1.0029 \\
$\alpha_\mathrm{xc}$   & 2539.2 & 1.0031 \\
$\alpha_\mathrm{w}$    & 2795.9 & 1.0030 \\
$\alpha_\mathrm{p}$    & 4871.6 & 1.0002 \\
$\alpha_\mathrm{r}$    & 2553.2 & 1.1926 \\
$\alpha_\mathrm{st}$   & 2563.7 & 1.0038 \\
$\alpha_\mathrm{m}$    & 3040.0 & 1.0002 \\
$\beta_\mathrm{m}$     & 2995.0 & 1.0305 \\
$\log\sigma$           & 5033.9 & 1.0002 \\
\bottomrule[1.0pt]
\end{tabular*}
\end{table}

\begin{table*}[htbp]
\caption{\label{tab:bwk_correlation}%
Pearson correlation matrix and Spearman rank correlation matrix of the posterior samples for the BWK* nuclear mass model parameters. See Sec.~\ref{sec:CorrelationDiagnosis}, Eqs.~(\ref{eq:Spearman}) and (\ref{eq:Pearson}) for the details.}
\begin{tabular*}{\linewidth}{@{\hspace{5mm}\extracolsep{\fill}}crrrrrrrrrrrr@{\hspace{5mm}}}
\toprule[1.0pt]
\midrule[0.25pt]
\multicolumn{13}{c}{Pearson correlation matrix}\\
 & $\alpha_\mathrm{v}$ & $\alpha_\mathrm{s}$ & $\alpha_\mathrm{c}$ & $\alpha_{\rm sym}$ & $\alpha_\mathrm{xc}$ & $\alpha_\mathrm{w}$ & $\alpha_\mathrm{p}$ & $\alpha_\mathrm{r}$ & $\alpha_\mathrm{st}$ & $\alpha_\mathrm{m}$ & $\beta_\mathrm{m}$ & $\log\sigma$ \\
\midrule[0.25pt]
$\alpha_\mathrm{v}$     & 1.000 & $-0.773$ & $-0.620$ & $-0.450$ & $-0.153$ & $-0.173$ & 0.034 & 0.814 & 0.163 & $-0.077$ & 0.074 & $-0.024$ \\
$\alpha_\mathrm{s}$     & $-0.773$ & 1.000 & 0.823 & 0.260 & $-0.462$ & $-0.123$ & 0.014 & $-0.911$ & $-0.214$ & 0.124 & $-0.142$ & 0.024 \\
$\alpha_\mathrm{c}$     & $-0.620$ & 0.823 & 1.000 & $-0.037$ & $-0.607$ & $-0.347$ & 0.020 & $-0.561$ & 0.076 & 0.247 & $-0.286$ & 0.016 \\
$\alpha_{\rm sym}$ & $-0.450$ & 0.260 & $-0.037$ & 1.000 & 0.323 & 0.692 & $-0.015$ & $-0.424$ & $-0.904$ & 0.076 & $-0.095$ & 0.021 \\
$\alpha_\mathrm{xc}$  & $-0.153$ & $-0.462$ & $-0.607$ & 0.323 & 1.000 & 0.559 & $-0.070$ & 0.152 & $-0.036$ & $-0.174$ & 0.219 & $-0.002$ \\
$\alpha_\mathrm{w}$     & $-0.173$ & $-0.123$ & $-0.347$ & 0.692 & 0.559 & 1.000 & $-0.038$ & $-0.175$ & $-0.664$ & $-0.022$ & 0.047 & 0.014 \\
$\alpha_\mathrm{p}$     & 0.034 & 0.014 & 0.020 & $-0.015$ & $-0.070$ & $-0.038$ & 1.000 & 0.006 & $-0.011$ & 0.061 & $-0.063$ & $-0.019$ \\
$\alpha_\mathrm{r}$     & 0.814 & $-0.911$ & $-0.561$ & $-0.424$ & 0.152 & $-0.175$ & 0.006 & 1.000 & 0.337 & $-0.072$ & 0.065 & $-0.027$ \\
$\alpha_\mathrm{st}$  & 0.163 & $-0.214$ & 0.076 & $-0.904$ & $-0.036$ & $-0.664$ & $-0.011$ & 0.337 & 1.000 & $-0.089$ & 0.110 & $-0.019$ \\
$\alpha_\mathrm{m}$     & $-0.077$ & 0.124 & 0.247 & 0.076 & $-0.174$ & $-0.022$ & 0.061 & $-0.072$ & $-0.089$ & 1.000 & $-0.920$ & 0.007 \\
$\beta_\mathrm{m}$      & 0.074 & $-0.142$ & $-0.286$ & $-0.095$ & 0.219 & 0.047 & $-0.063$ & 0.065 & 0.110 & $-0.920$ & 1.000 & $-0.005$ \\
$\log\sigma$   & $-0.024$ & 0.024 & 0.016 & 0.021 & $-0.002$ & 0.014 & $-0.019$ & $-0.027$ & $-0.019$ & 0.007 & $-0.005$ & 1.000 \\

\toprule[1.0pt]
\midrule[0.25pt]
\multicolumn{13}{c}{Spearman rank correlation matrix}\\
 & $\alpha_\mathrm{v}$ & $\alpha_\mathrm{s}$ & $\alpha_\mathrm{c}$ & $\alpha_{\rm sym}$ & $\alpha_\mathrm{xc}$ & $\alpha_\mathrm{w}$ & $\alpha_\mathrm{p}$ & $\alpha_\mathrm{r}$ & $\alpha_\mathrm{st}$ & $\alpha_\mathrm{m}$ & $\beta_\mathrm{m}$ & $\log\sigma$ \\
\midrule[0.25pt]
$\alpha_\mathrm{v}$ & 1.000 & $-0.724$ & $-0.580$ & $-0.452$ & $-0.159$ & $-0.175$ & 0.028 & 0.782 & 0.153 & $-0.069$ & 0.065 & $-0.020$ \\
$\alpha_\mathrm{s}$ & $-0.724$ & 1.000 & 0.816 & 0.226 & $-0.478$ & $-0.143$ & 0.019 & $-0.886$ & $-0.177$ & 0.104 & $-0.121$ & 0.015 \\
$\alpha_\mathrm{c}$ & $-0.580$ & 0.816 & 1.000 & $-0.037$ & $-0.598$ & $-0.342$ & 0.021 & $-0.530$ & 0.082 & 0.224 & $-0.261$ & 0.009 \\
$\alpha_{\rm sym}$ & $-0.452$ & 0.226 & $-0.037$ & 1.000 & 0.289 & 0.660 & $-0.012$ & $-0.405$ & $-0.886$ & 0.098 & $-0.113$ & 0.021 \\
$\alpha_\mathrm{xc}$ & $-0.159$ & $-0.478$ & $-0.598$ & 0.289 & 1.000 & 0.531 & $-0.067$ & 0.156 & $-0.001$ & $-0.155$ & 0.200 & $-0.003$ \\
$\alpha_\mathrm{w}$ & $-0.175$ & $-0.143$ & $-0.342$ & 0.660 & 0.531 & 1.000 & $-0.033$ & $-0.168$ & $-0.632$ & $-0.005$ & 0.033 & 0.011 \\
$\alpha_\mathrm{p}$ & 0.028 & 0.019 & 0.021 & $-0.012$ & $-0.067$ & $-0.033$ & 1.000 & 0.002 & $-0.012$ & 0.053 & $-0.058$ & $-0.017$ \\
$\alpha_\mathrm{r}$ & 0.782 & $-0.886$ & $-0.530$ & $-0.405$ & 0.156 & $-0.168$ & 0.002 & 1.000 & 0.311 & $-0.048$ & 0.041 & $-0.022$ \\
$\alpha_\mathrm{st}$ & 0.153 & $-0.177$ & 0.082 & $-0.886$ & $-0.001$ & $-0.632$ & $-0.012$ & 0.311 & 1.000 & $-0.103$ & 0.121 & $-0.021$ \\
$\alpha_\mathrm{m}$ & $-0.069$ & 0.104 & 0.224 & 0.098 & $-0.155$ & $-0.005$ & 0.053 & $-0.048$ & $-0.103$ & 1.000 & $-0.913$ & 0.010 \\
$\beta_\mathrm{m}$ & 0.065 & $-0.121$ & $-0.261$ & $-0.113$ & 0.200 & 0.033 & $-0.058$ & 0.041 & 0.121 & $-0.913$ & 1.000 & $-0.008$ \\
$\log\sigma$ & $-0.020$ & 0.015 & 0.009 & 0.021 & $-0.003$ & 0.011 & $-0.017$ & $-0.022$ & $-0.021$ & 0.010 & $-0.008$ & 1.000 \\
\bottomrule[1.0pt]
\end{tabular*}
\end{table*}

\subsection{BWK mass model}
\label{sec:BWK_MCMC}

\begin{figure*}[t]
\includegraphics[width=\textwidth]{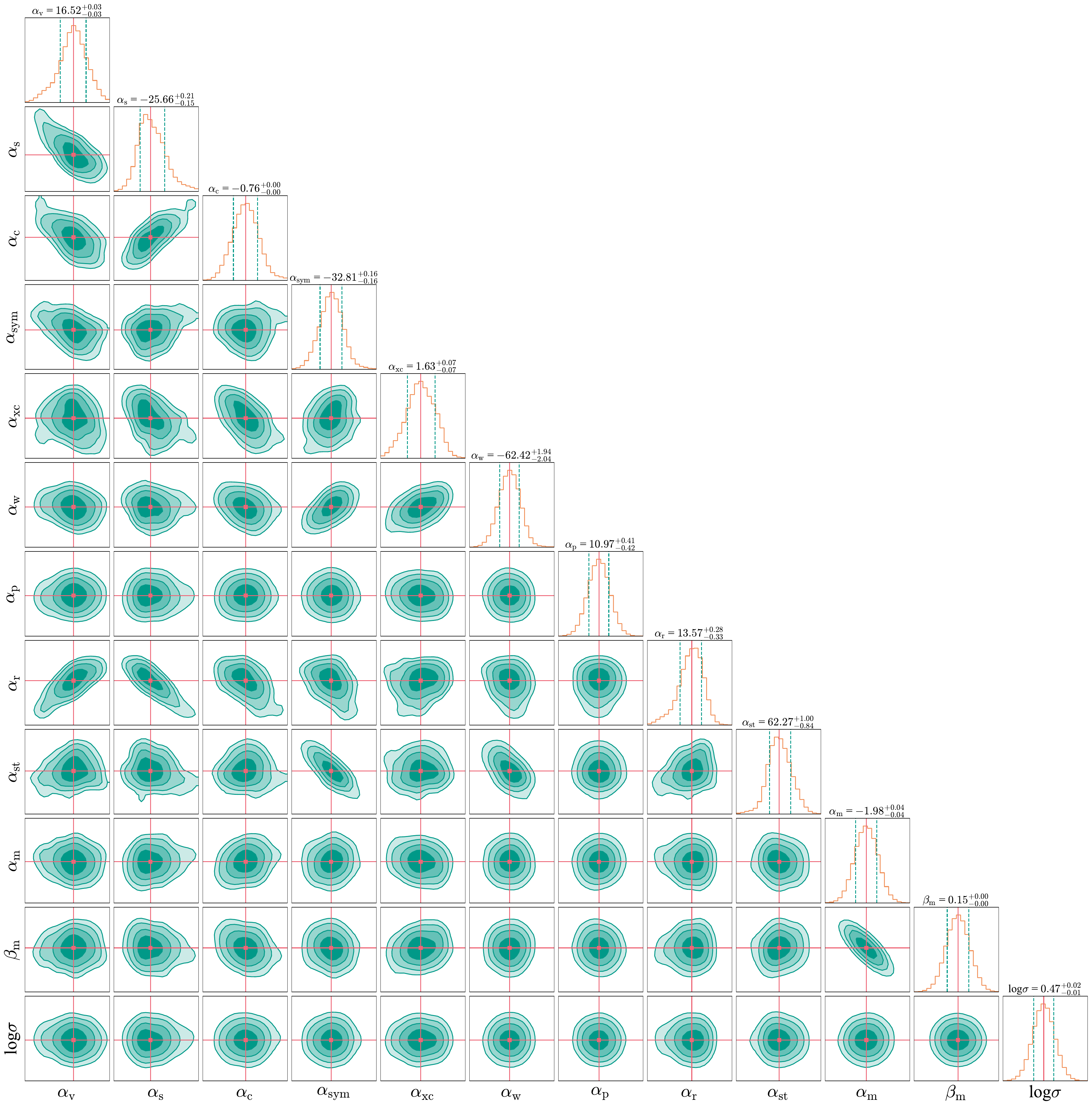}%
\makebox[0pt][r]{%
\raisebox{40em}{%
\includegraphics[width=0.6\linewidth]{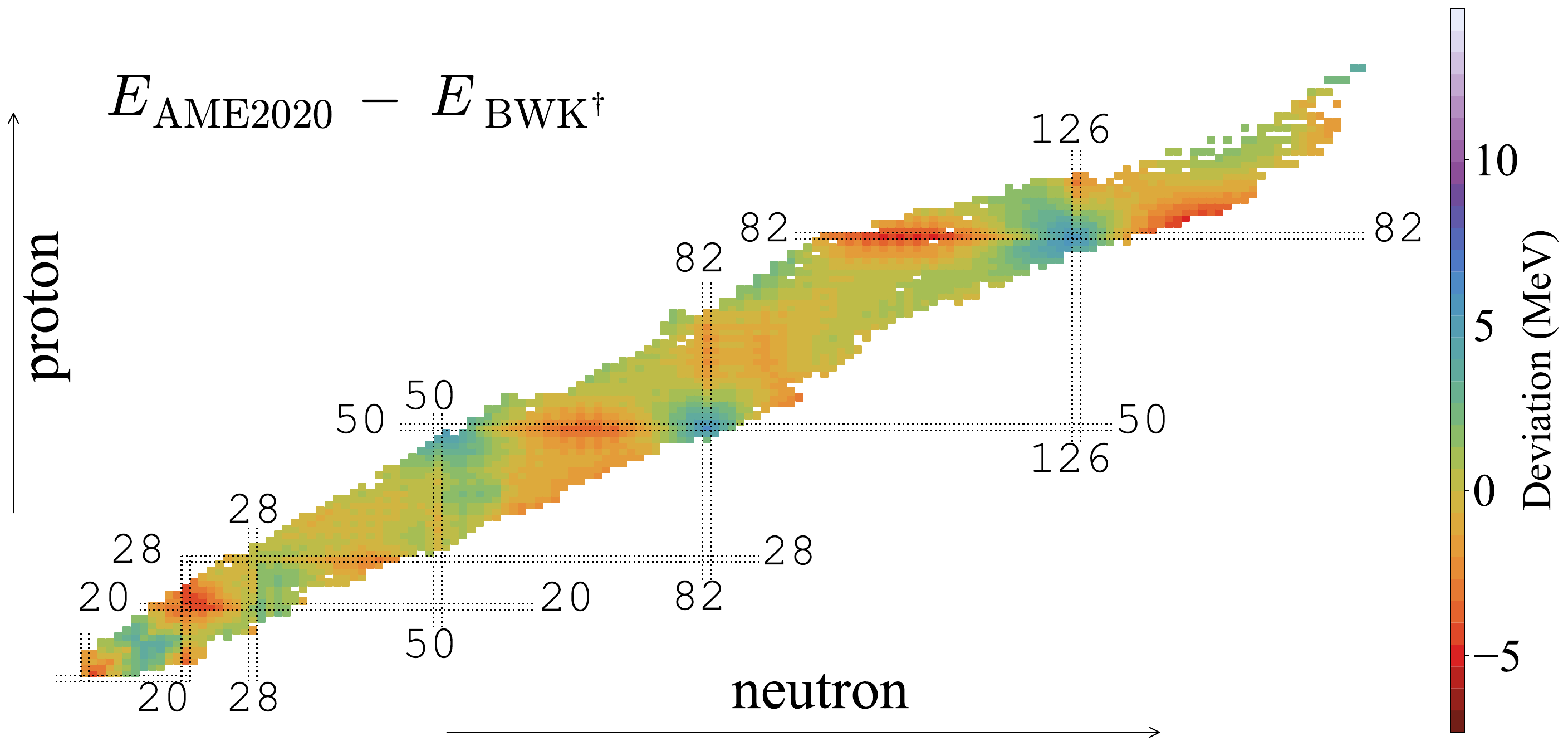}%
}\hspace*{-1em}%
}%
\caption{\label{fig:posterior_BWK}Posterior distributions for all 11 parameters of the BWK* nuclear mass model. The most probable distribution of each parameter of BWK* is shown on top of the respective diagonal plot. The mean values of posterior distributions of all 11 parameters along with the corresponding physical units of BWK* are listed in the sixth column of Table~\ref{tab:mass_models_01}. The optimal parameter set is presented in the seventh column of Table~\ref{tab:mass_models_01}. The deviation of BWK$^\dagger$ and AME2020 binding energies is presented at the nuclear chart with respect to the right color scale. See Fig.~\ref{fig:posterior_BW} for the rest of description. }
\end{figure*}

Table~\ref{tab:mass_model_BWK} summarizes the effective sample size, $n_\mathrm{eff}$, and Gelman-Rubin statistics, $\hat{R}$, for all parameters of the BWK* mass model. All $\hat{R}$ values below 1.04 indicate a convergence of Markov chain. The noticeably $\hat{R}=1.1926$ for $\alpha_\mathrm{r}$ exposes a local Markov chain mixing bottleneck, indicating a statistical signature of parameter degeneracies and strength absorption by other parameters. All these $n_\mathrm{eff}$ values exceed 2500, ensuring reliable posterior estimates. The posterior mean of the noise standard deviation $\sigma$ is 1.601~MeV with a $95~\%$ credible interval of  [1.55, 1.65]~MeV.

The posterior distributions of the BWK* model show that the posterior distributions of several parameters deviate noticeably from Gaussian forms (diagonal plots in Fig.~\ref{fig:posterior_BWK}) with mild skewness and heavier tails, which are not found in the BW* model. This reflects the nonlinearity introduced by the additional terms in the BWK* model, increasing the complexity of the posterior landscape, which could not be captured by least-squares fit methods. Moreover, negative nonlinearity but monotonic correlations among several key parameters are also revealed in Table~\ref{tab:bwk_correlation}, presenting the Pearson correlation and Spearman rank correlation matrices of the BWK* model, e.g., 
$\rho(\alpha_\mathrm{sym},\alpha_\mathrm{st}) = -0.886$, 
$\rho(\alpha_\mathrm{w},\alpha_\mathrm{st}) = -0.632$, and 
$\rho(\alpha_\mathrm{m},\beta_\mathrm{m}) = -0.913$. 
Positive correlations are also observed, e.g., $\rho(\alpha_\mathrm{sym},\alpha_\mathrm{w}) = 0.660$. These results agree with the finding of \citet{Kirson2008} that the interplay and compensatory effects between different terms are important features of the semi-empirical mass formula.

\subsection{BWN mass model}
\label{sec:BWN_MCMC}

\begin{figure*}
\includegraphics[width=\textwidth]{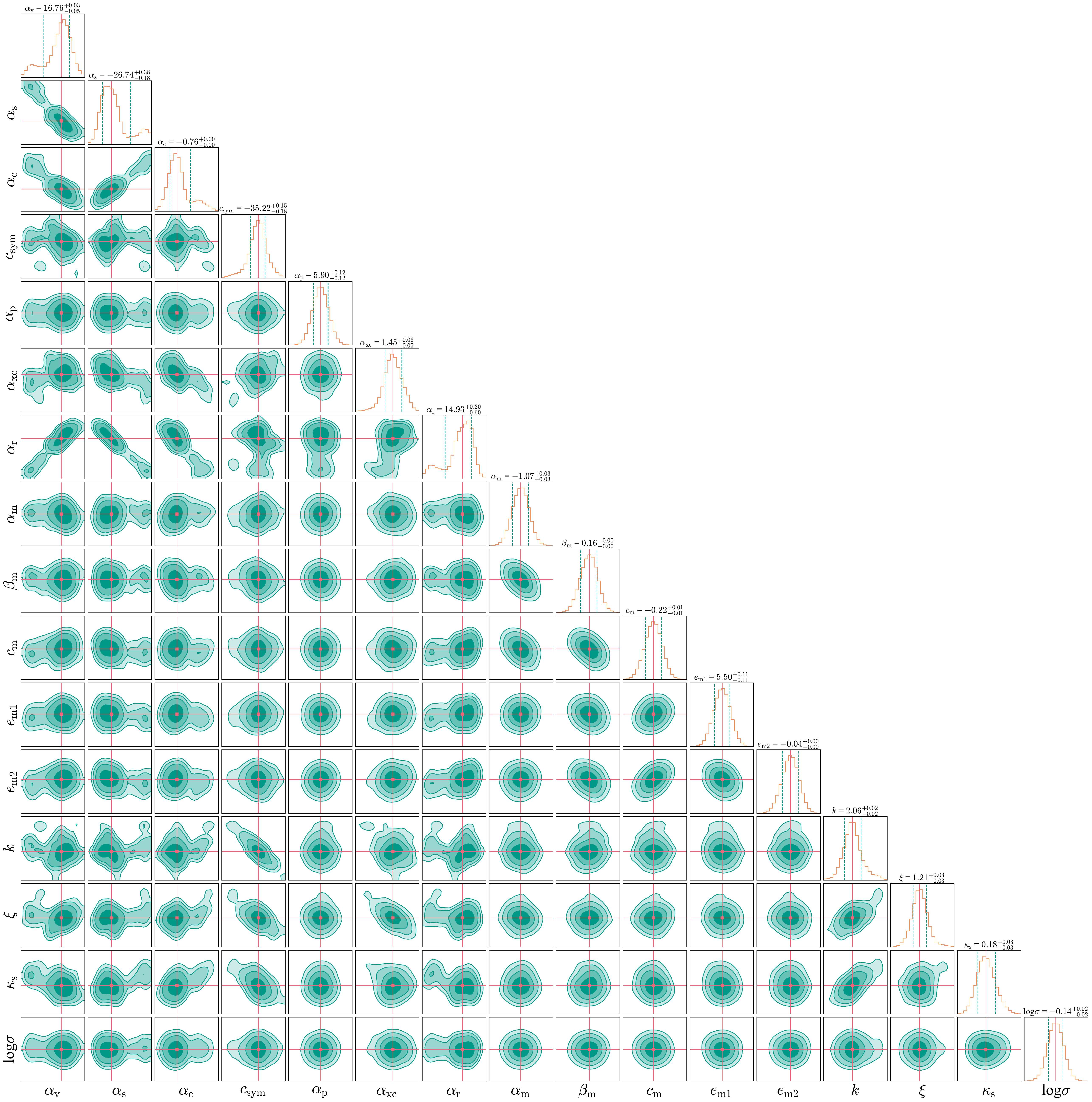}%
\makebox[0pt][r]{%
\raisebox{40em}{%
\includegraphics[width=0.6\linewidth]{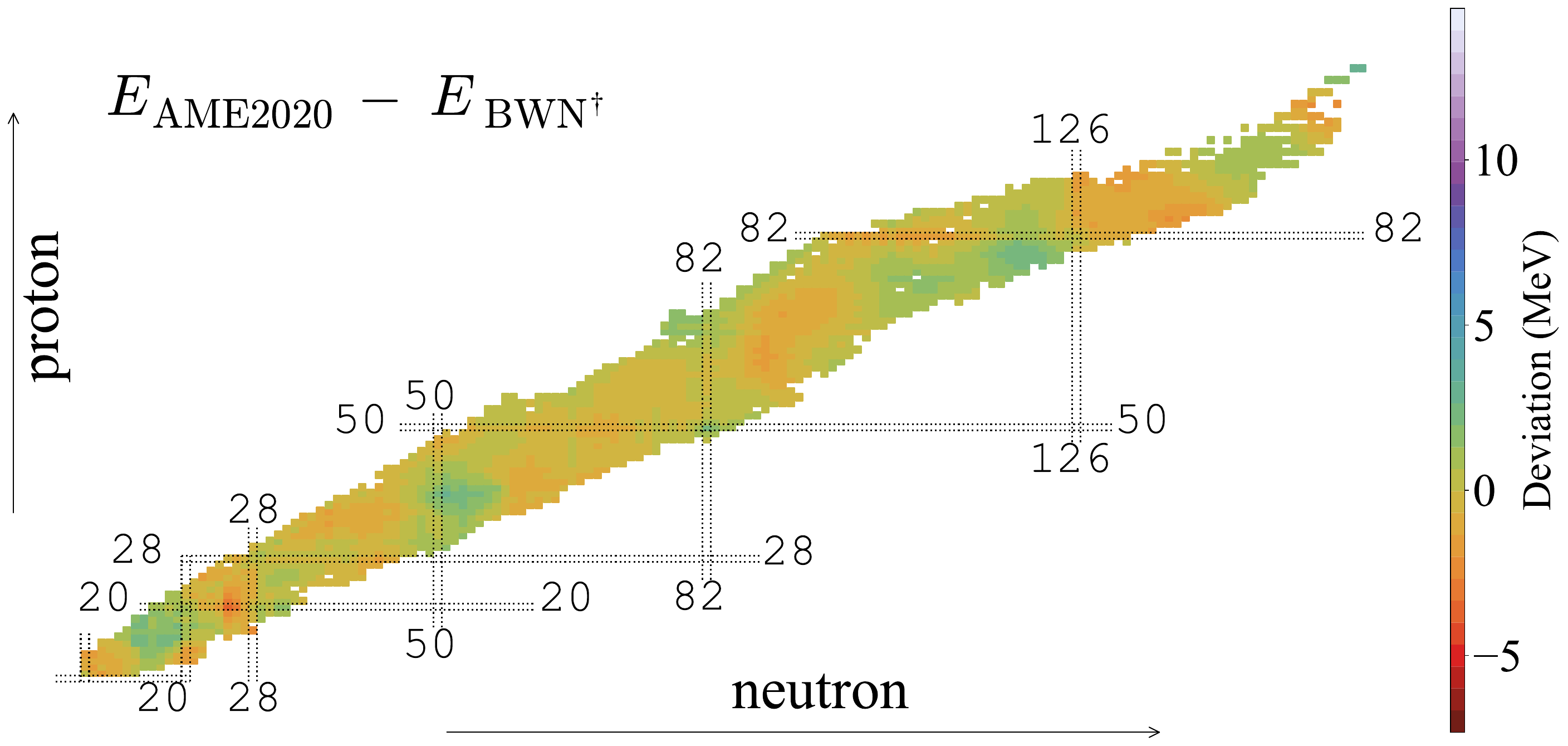}%
}\hspace*{-1em}%
}%
\caption{\label{fig:posterior_BWN}Posterior distributions for all 15 parameters of the BWN* nuclear mass model. The most probable distribution of each parameter of BWN* is shown on top of the respective diagonal plot. The mean values of posterior distributions of all 15 parameters along with the corresponding physical units of BWN* are listed in the third column of Table~\ref{tab:mass_models_02}. The optimal parameter set is presented in the fourth column of Table~\ref{tab:mass_models_02}. The deviation of BWN$^\dagger$ and AME2020 binding energies is presented at the nuclear chart with respect to the right color scale. See Fig.~\ref{fig:posterior_BW} for the rest of description.}
\end{figure*}

\begin{table}[t!]
\caption{\label{tab:mass_model_BWN}%
Convergence diagnosis for the reference chain of BWN*. See Sec.~\ref{sec:MultiChain}, Eqs.~(\ref{eq:GelmanRubin}) and (\ref{eq:neff}) for the details.}
\begin{tabular*}{\linewidth}{@{\hspace{10mm}\extracolsep{\fill}}ccc@{\hspace{10mm}}}
\toprule[1.0pt]
\midrule[0.25pt]Parameter & $n_\mathrm{eff}$
 & $\hat{R}$ \\
\midrule[0.25pt]
$\alpha_\mathrm{v}$    & 2505.5 & 1.0001  \\
$\alpha_\mathrm{s}$    & 2503.9 & 1.0010 \\
$\alpha_\mathrm{c}$   & 2514.7 &  1.0001 \\
$c_\mathrm{sym}$  & 2520.4 &   1.0001    \\
$\alpha_\mathrm{xc}$  & 2520.1 & 1.0232  \\
$\alpha_\mathrm{p}$ & 3926.0 & 1.0001 \\
$\alpha_\mathrm{r}$ & 2507.9 & 1.0289 \\
$\alpha_\mathrm{m}$ & 2995.2 &  1.0006 \\
$\beta_\mathrm{m}$ & 2845.5 & 1.0350 \\
$c_\mathrm{m}$ & 2961.6 & 1.0025 \\
$e_\mathrm{m1}$ & 3600.8 & 1.0019 \\
$e_\mathrm{m2}$ & 3210.9 & 1.0024 \\
$k$ & 2521.4 & 1.0011 \\
$\xi$ & 2621.6 & 1.0317 \\
$\kappa_\mathrm{s}$ & 2753.4 & 1.0285 \\
$\log\sigma$       & 4672.6  &  1.0000 \\

\bottomrule[1.0pt]
\end{tabular*}
\end{table}

\begin{sidewaystable*}[htbp]
\vspace{70mm}
\caption{\label{tab:bwn_correlation}%
Pearson correlation matrix and Spearman rank correlation matrix of the posterior samples for the BWN* nuclear mass model parameters. See Sec.~\ref{sec:CorrelationDiagnosis}, Eqs.~(\ref{eq:Spearman}) and (\ref{eq:Pearson}) for the details.}
\begin{tabular*}{\linewidth}{@{\hspace{5mm}\extracolsep{\fill}}crrrrrrrrrrrrrrrr@{\hspace{5mm}}}
\toprule[1.0pt]
\midrule[0.25pt]
\multicolumn{17}{c}{Pearson correlation matrix}\\
 & $\alpha_\mathrm{v}$ & $\alpha_\mathrm{s}$ & $\alpha_\mathrm{c}$ & $c_{\rm sym}$ & $\alpha_\mathrm{p}$ & $\alpha_\mathrm{xc}$ & $\alpha_\mathrm{r}$ & $\alpha_\mathrm{m}$ & $\beta_\mathrm{m}$ & $c_\mathrm{m}$ & $e_\mathrm{m1}$ & $e_\mathrm{m2}$ & $k$ & $\xi$ & $\kappa_\mathrm{s}$ & $\log\sigma$ \\
\midrule[0.25pt]
$\alpha_\mathrm{v}$     & 1.000 & $-0.927$ & $-0.883$ & $-0.190$ & 0.029 & 0.275 & 0.938 & $-0.133$ & 0.012 & 0.202 & 0.148 & 0.177 & $-0.098$ & $-0.024$ & $-0.464$ & $-0.056$ \\
$\alpha_\mathrm{s}$     & $-0.927$ & 1.000 & 0.936 & 0.077 & $-0.014$ & $-0.591$ & $-0.982$ & 0.103 & 0.017 & $-0.225$ & $-0.179$ & $-0.238$ & 0.110 & 0.268 & 0.399 & 0.057 \\
$\alpha_\mathrm{c}$     & $-0.883$ & 0.936 & 1.000 & $-0.108$ & $-0.037$ & $-0.642$ & $-0.869$ & 0.135 & $-0.122$ & $-0.081$ & $-0.109$ & $-0.070$ & 0.293 & 0.353 & 0.532 & 0.058 \\
$c_{\rm sym}$  & $-0.190$ & 0.077 & $-0.108$ & 1.000 & $-0.019$ & 0.313 & $-0.157$ & 0.107 & $-0.122$ & $-0.013$ & 0.041 & $-0.023$ & $-0.924$ & $-0.673$ & $-0.579$ & 0.006 \\
$\alpha_\mathrm{p}$     & 0.029 & $-0.014$ & $-0.037$ & $-0.019$ & 1.000 & 0.013 & $-0.024$ & 0.036 & $-0.047$ & 0.014 & $-0.000$ & 0.048 & 0.008 & $-0.003$ & $-0.012$ & 0.013 \\
$\alpha_\mathrm{xc}$  & 0.275 & $-0.591$ & $-0.642$ & 0.313 & 0.013 & 1.000 & 0.471 & $-0.019$ & 0.093 & $-0.015$ & 0.072 & 0.057 & $-0.200$ & $-0.700$ & $-0.149$ & $-0.031$ \\
$\alpha_\mathrm{r}$     & 0.938 & $-0.982$ & $-0.869$ & $-0.157$ & $-0.024$ & 0.471 & 1.000 & $-0.105$ & $-0.069$ & 0.288 & 0.206 & 0.299 & $-0.046$ & $-0.155$ & $-0.370$ & $-0.054$ \\
$\alpha_\mathrm{m}$     & $-0.133$ & 0.103 & 0.135 & 0.107 & 0.036 & $-0.019$ & $-0.105$ & 1.000 & $-0.548$ & $-0.402$ & 0.044 & 0.026 & $-0.083$ & $-0.114$ & 0.009 & 0.001 \\
$\beta_\mathrm{m}$      & 0.012 & 0.017 & $-0.122$ & $-0.122$ & $-0.047$ & 0.093 & $-0.069$ & $-0.548$ & 1.000 & $-0.472$ & $-0.204$ & $-0.321$ & 0.095 & 0.003 & 0.086 & 0.006 \\
$c_\mathrm{m}$          & 0.202 & $-0.225$ & $-0.081$ & $-0.013$ & 0.014 & $-0.015$ & 0.288 & $-0.402$ & $-0.472$ & 1.000 & 0.228 & 0.462 & 0.004 & 0.080 & $-0.112$ & $-0.011$ \\
$e_\mathrm{m1}$       & 0.148 & $-0.179$ & $-0.109$ & 0.041 & $-0.000$ & 0.072 & 0.206 & 0.044 & $-0.204$ & 0.228 & 1.000 & $-0.264$ & $-0.072$ & $-0.069$ & $-0.068$ & $-0.028$ \\
$e_\mathrm{m2}$       & 0.177 & $-0.238$ & $-0.070$ & $-0.023$ & 0.048 & 0.057 & 0.299 & 0.026 & $-0.321$ & 0.462 & $-0.264$ & 1.000 & 0.002 & $-0.076$ & $-0.038$ & 0.010 \\
$k$            & $-0.098$ & 0.110 & 0.293 & $-0.924$ & 0.008 & $-0.200$ & $-0.046$ & $-0.083$ & 0.095 & 0.004 & $-0.072$ & 0.002 & 1.000 & 0.632 & 0.712 & 0.008 \\
$\xi$          & $-0.024$ & 0.268 & 0.353 & $-0.673$ & $-0.003$ & $-0.700$ & $-0.155$ & $-0.114$ & 0.003 & 0.080 & $-0.069$ & $-0.076$ & 0.632 & 1.000 & 0.195 & 0.013 \\
$\kappa_\mathrm{s}$     & $-0.464$ & 0.399 & 0.532 & $-0.579$ & $-0.012$ & $-0.149$ & $-0.370$ & 0.009 & 0.086 & $-0.112$ & $-0.068$ & $-0.038$ & 0.712 & 0.195 & 1.000 & 0.025 \\
$\log\sigma$   & $-0.056$ & 0.057 & 0.058 & 0.006 & 0.013 & $-0.031$ & $-0.054$ & 0.001 & 0.006 & $-0.011$ & $-0.028$ & 0.010 & 0.008 & 0.013 & 0.025 & 1.000 \\

\toprule[1.0pt]
\midrule[0.25pt]
\multicolumn{17}{c}{Spearman rank correlation matrix}\\
 & $\alpha_\mathrm{v}$ & $\alpha_\mathrm{s}$ & $\alpha_\mathrm{c}$ & $c_{\rm sym}$ & $\alpha_\mathrm{p}$ & $\alpha_\mathrm{xc}$ & $\alpha_\mathrm{r}$ & $\alpha_\mathrm{m}$ & $\beta_\mathrm{m}$ & $c_\mathrm{m}$ & $e_\mathrm{m1}$ & $e_\mathrm{m2}$ & $k$ & $\xi$ & $\kappa_\mathrm{s}$ & $\log\sigma$ \\
\midrule[0.25pt]
$\alpha_\mathrm{v}$ & 1.000 & $-0.870$ & $-0.801$ & $-0.308$ & 0.028 & 0.148 & 0.886 & $-0.166$ & 0.032 & 0.191 & 0.117 & 0.149 & $-0.016$ & 0.104 & $-0.397$ & $-0.039$ \\
$\alpha_\mathrm{s}$ & $-0.870$ & 1.000 & 0.862 & 0.258 & $-0.015$ & $-0.529$ & $-0.968$ & 0.122 & 0.009 & $-0.223$ & $-0.152$ & $-0.232$ & $-0.063$ & 0.139 & 0.255 & 0.036 \\
$\alpha_\mathrm{c}$ & $-0.801$ & 0.862 & 1.000 & 0.027 & $-0.046$ & $-0.580$ & $-0.750$ & 0.171 & $-0.167$ & $-0.047$ & $-0.052$ & $-0.026$ & 0.177 & 0.234 & 0.443 & 0.044 \\
$c_{\rm sym}$ & $-0.308$ & 0.258 & 0.027 & 1.000 & $-0.008$ & 0.157 & $-0.311$ & 0.103 & $-0.132$ & $-0.005$ & 0.034 & $-0.025$ & $-0.881$ & $-0.556$ & $-0.524$ & 0.009 \\
$\alpha_\mathrm{p}$ & 0.028 & $-0.015$ & $-0.046$ & $-0.008$ & 1.000 & 0.021 & $-0.032$ & 0.030 & $-0.041$ & 0.016 & $-0.001$ & 0.057 & $-0.004$ & $-0.022$ & $-0.014$ & 0.017 \\
$\alpha_\mathrm{xc}$ & 0.148 & $-0.529$ & $-0.580$ & 0.157 & 0.021 & 1.000 & 0.380 & $-0.021$ & 0.094 & $-0.020$ & 0.051 & 0.058 & $-0.043$ & $-0.612$ & $-0.048$ & $-0.018$ \\
$\alpha_\mathrm{r}$ & 0.886 & $-0.968$ & $-0.750$ & $-0.311$ & $-0.032$ & 0.380 & 1.000 & $-0.122$ & $-0.076$ & 0.299 & 0.192 & 0.301 & 0.095 & $-0.021$ & $-0.249$ & $-0.033$ \\
$\alpha_\mathrm{m}$ & $-0.166$ & 0.122 & 0.171 & 0.103 & 0.030 & $-0.021$ & $-0.122$ & 1.000 & $-0.523$ & $-0.380$ & 0.030 & 0.026 & $-0.060$ & $-0.122$ & 0.019 & $-0.001$ \\
$\beta_\mathrm{m}$ & 0.032 & 0.009 & $-0.167$ & $-0.132$ & $-0.041$ & 0.094 & $-0.076$ & $-0.523$ & 1.000 & $-0.474$ & $-0.202$ & $-0.323$ & 0.096 & 0.012 & 0.095 & 0.010 \\
$c_\mathrm{m}$ & 0.191 & $-0.223$ & $-0.047$ & $-0.005$ & 0.016 & $-0.020$ & 0.299 & $-0.380$ & $-0.474$ & 1.000 & 0.234 & 0.452 & $-0.012$ & 0.084 & $-0.122$ & $-0.009$ \\
$e_\mathrm{m1}$ & 0.117 & $-0.152$ & $-0.052$ & 0.034 & $-0.001$ & 0.051 & 0.192 & 0.030 & $-0.202$ & 0.234 & 1.000 & $-0.246$ & $-0.072$ & $-0.065$ & $-0.069$ & $-0.021$ \\
$e_\mathrm{m2}$ & 0.149 & $-0.232$ & $-0.026$ & $-0.025$ & 0.057 & 0.058 & 0.301 & 0.026 & $-0.323$ & 0.452 & $-0.246$ & 1.000 & 0.000 & $-0.074$ & $-0.041$ & 0.009 \\
$k$ & $-0.016$ & $-0.063$ & 0.177 & $-0.881$ & $-0.004$ & $-0.043$ & 0.095 & $-0.060$ & 0.096 & $-0.012$ & $-0.072$ & 0.000 & 1.000 & 0.520 & 0.685 & 0.010 \\
$\xi$ & 0.104 & 0.139 & 0.234 & $-0.556$ & $-0.022$ & $-0.612$ & $-0.021$ & $-0.122$ & 0.012 & 0.084 & $-0.065$ & $-0.074$ & 0.520 & 1.000 & 0.100 & 0.005 \\
$\kappa_\mathrm{s}$ & $-0.397$ & 0.255 & 0.443 & $-0.524$ & $-0.014$ & $-0.048$ & $-0.249$ & 0.019 & 0.095 & $-0.122$ & $-0.069$ & $-0.041$ & 0.685 & 0.100 & 1.000 & 0.021 \\
$\log\sigma$ & $-0.039$ & 0.036 & 0.044 & 0.009 & 0.017 & $-0.018$ & $-0.033$ & $-0.001$ & 0.010 & $-0.009$ & $-0.021$ & 0.009 & 0.010 & 0.005 & 0.021 & 1.000 \\
\bottomrule[1.0pt]
\end{tabular*}%
\end{sidewaystable*}

The analysis of posterior distributions reveals mild multi-modality in the posterior distributions of several BWN* parameters, i.e., $ \alpha_\mathrm{v} $, $ \alpha_\mathrm{s} $, $ \alpha_\mathrm{c} $, and $\alpha_\mathrm{r}$ (diagonal plots in Fig.~\ref{fig:posterior_BWN}). Table~\ref{tab:mass_model_BWN} shows that the Gelman-Rubin statistics $  \hat{R} $ of all parameters remain below 1.04 and $n_\mathrm{eff}$ exceed 2500 for all parameters, indicating that the multi-chain MCMC sampling for the BWN* model achieves excellent convergence. This multi-modality is not a sampling artifact, but rather reflects the structures arising from the shell-correction and pairing terms introduced in the BWN* model (Eq.~(\ref{eq:BWN})).

\newpage
The main non-symmetric correlations of 
$\alpha_\mathrm{sym}$-$\alpha_\mathrm{st}$, 
$\alpha_\mathrm{w}$-$\alpha_\mathrm{st}$,
$\alpha_\mathrm{m}$-$\beta_\mathrm{m}$, and 
$\alpha_\mathrm{sym}$-$\alpha_\mathrm{w}$ 
in the BWK* model are inherited by the BWN* model. The detailed correlation strengths among the model parameters are presented in Table~\ref{tab:bwn_correlation}. The symmetry energy term $\alpha_{\rm sym}$ and the related Wigner and surface symmetry correction terms in the BWK model are folded to a single term with the redefined $\alpha_{\rm sym}$ consisting of $c_{\rm sym}$, $k$, and $\xi$, and with the surface diffuseness factor $f_s$ calibrated by $\kappa_\mathrm{s}$ in the BWN* model. The original BW correlation in the symmetry term has been restructured, establishing a set of separable non-linear terms, i.e., (i) $k/A^{1/3}$ term, accounting for the reduction of symmetry energy due to the surface effects, (ii) $\xi$ term, providing {\it ad hoc} corrections for nuclei with extreme isospin asymmetry, and (iii) $f_\mathrm{s}$ factor, introducing the contribution of surface diffuseness, which implicitly correlates with the nuclear surface properties. The former two separable terms fine tune the BWN model to better reproduce experimental binding energies in regions far from the $\beta$-stability line than the previous linear form.

According to the Spearman rank correlation matrix in Table~\ref{tab:bwn_correlation}, the parameters of new terms in the BWN* model show non-linear yet monotonic correlations, i.e., $ \rho(c_{\rm sym},\xi) = -0.556 $, $ \rho(c_{\rm sym},k) = -0.881 $, $ \rho(c_{\rm sym},\kappa_\mathrm{s}) = -0.524 $, $ \rho(k,\xi) = 0.520 $, and $ \rho(k,\kappa_\mathrm{s}) = 0.685 $. Meanwhile, the posterior distributions in the BWN* model exhibit multi-modality (diagonal plots of $\alpha_\mathrm{v}$, $\alpha_\mathrm{s}$, $\alpha_\mathrm{c}$, and $\alpha_\mathrm{r}$ in Fig.~\ref{fig:posterior_BWN}) and parameter degeneracies (e.g., covariance plots of $\alpha_\mathrm{v}$-$c_{\rm sym}$, $c_{\rm sym}$-$\alpha_\mathrm{r}$, $\alpha_\mathrm{v}$-$k$ and $\alpha_\mathrm{xc}$-$k$ in Fig.~\ref{fig:posterior_BWN}).

By using the condition number of Pearson correlation matrix to analyze the parameter degeneracy of the BWN* model, we obtain $\mathcal{K}(\mathbf{R}_\mathrm{P}) = 1.42 \times 10^6$, showing a high linear degeneracy. However, this outcome contradicts to the covariance plots in Fig.~\ref{fig:posterior_BWN}. For the BWN* model, the condition number of the Spearman rank correlation matrix, $\mathcal{K}(\mathbf{R}_\mathrm{S}) = 8.14 \times 10^2 < 10^3$ (Eq.~(\ref{eq:condition_number})), which implies that the degeneracy is rather weak, corresponding to the overall covariance plots in Fig~\ref{fig:posterior_BWN}. The compensation effects could originate from non-linear parameter manifolds rather than linear degeneracy. The weak parameter degeneracies could be solved by introducing new terms or deformation corrections to reshape the posterior parameter space. In this work, we take into account deformation corrections (Sec.~\ref{sec:BWL_MCMC}) for alleviating the parameter degeneracies, whereas the introduction of new term(s) could be future potential work and beyond the scope of this work.

The corner plots in Fig.~\ref{fig:posterior_BWN} reveal negative correlations among $c_\mathrm{m}$, $\alpha_\mathrm{m}$, and $\beta_\mathrm{m}$. The Spearman rank correlation matrix elements of $c_\mathrm{m}$-$\alpha_\mathrm{m}$, $c_\mathrm{m}$-$\beta_\mathrm{m}$, and $\alpha_\mathrm{m}$-$\beta_\mathrm{m}$ in Table~\ref{tab:bwn_correlation} indicate the numerical details of these negative correlations. The \(c_\mathrm{m}(v_\mathrm{n} + v_\mathrm{p})\) term is a linear correction with \(c_\mathrm{m} = -0.22\pm0.01\) (posterior plot of $c_\mathrm{m}$ in Fig.~\ref{fig:posterior_BWN}), decreasing the influence incurred from shell effect as the total valence number increases away from magic numbers. For regions of large valence numbers, this linear term is interrelated with \(\alpha_\mathrm{m} P\), such that when the \(c_\mathrm{m}(v_\mathrm{n} + v_\mathrm{p})\) term becomes more negative, \(\alpha_\mathrm{m} P\) tends to maintain the overall shell correction energy. Similarly, \(c_\mathrm{m}\) is interrelated with \(\beta_\mathrm{m}\) because the cumulative effect of the \(c_\mathrm{m}(v_\mathrm{n} + v_\mathrm{p})\) term partially absorbs the influence of the quadratic \(\beta_\mathrm{m} P^2\) term for the regions between magic numbers.

In addition, a weak positive correlation is observed between \(c_\mathrm{m}\) and \(e_\mathrm{m2}\). The exponential term of \(e_\mathrm{m1} \delta_{\rm shell} \exp(e_\mathrm{m2} (v_\mathrm{p}^2 + v_\mathrm{n}^2))\) (with \(e_\mathrm{m2} < 0\)) provides a strong enhanced (quenched) shell correction near (away from) magic numbers. The mutual balance between the exponential term and the \(c_\mathrm{m}(v_\mathrm{n} + v_\mathrm{p})\) linear term is established, of which the exponential term dominates the enhancement in regions near closed shells, while the linear term smoothly quenches the correction for large valence numbers. For instance, for regions near magic numbers, with less negative \(e_\mathrm{m2}\) and thus \(c_\mathrm{m}\) also having to be less negative, the shell effect is strengthened, demonstrating the compensatory nature. 
The analysis of BA-MCMC naturally indicates this behavior, consistent with the least-squares fit results presented in Ref.~\cite{Wu2025}. See Table~\ref{tab:mass_models_02} for (\(e_\mathrm{m2} = -0.0444\), \(c_\mathrm{m} = -0.2343\)) of BWN obtained from least-squares fit and (\(e_\mathrm{m2} = -0.0420\), \(c_\mathrm{m} = -0.2243\)) of BWN* generated from the BA-MCMC.

\subsection{BWL mass model}
\label{sec:BWL_MCMC}

\begin{figure*}
\includegraphics[width=\textwidth]{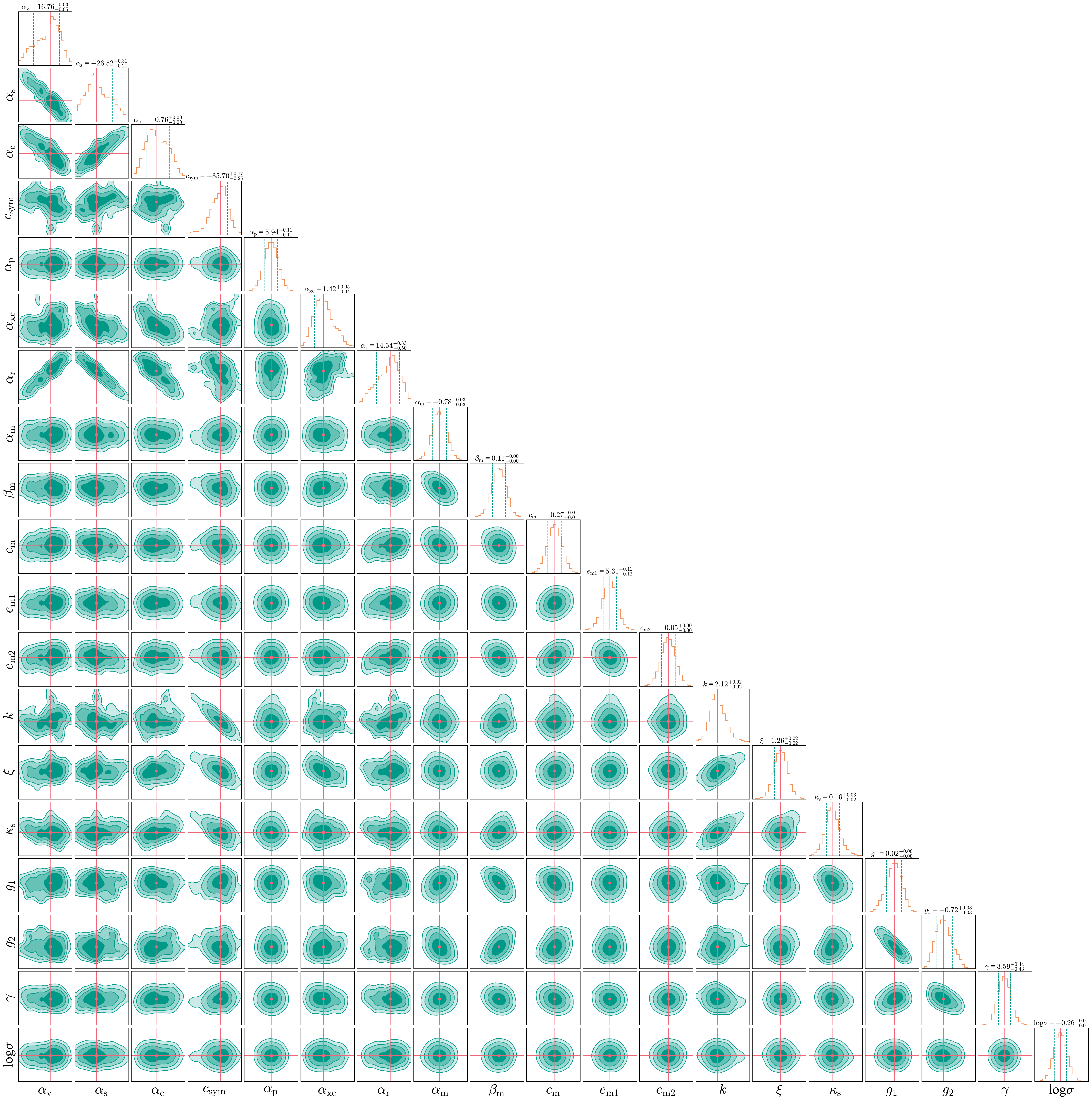}%
\makebox[0pt][r]{%
\raisebox{40em}{%
\includegraphics[width=0.6\linewidth]{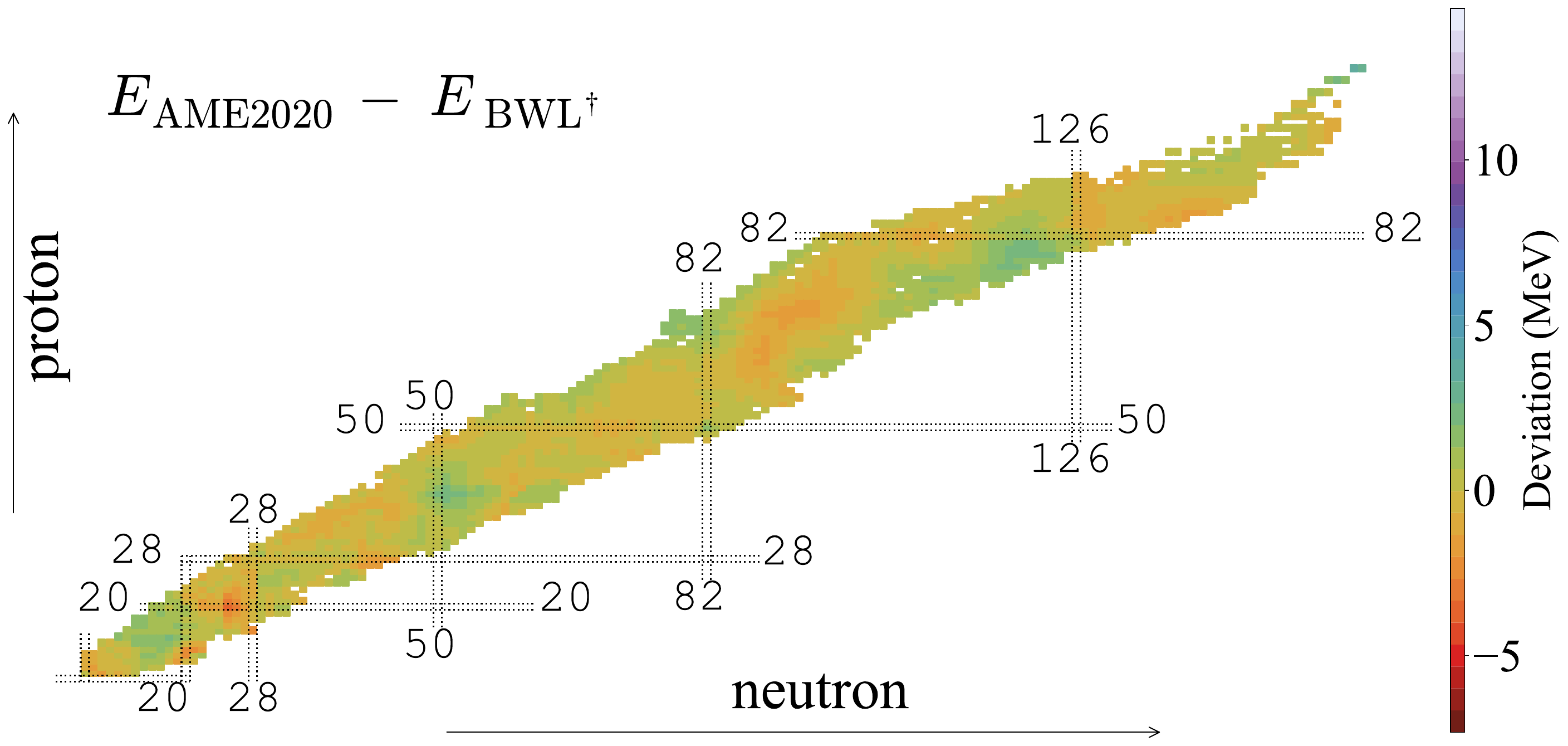}%
}\hspace*{-1em}%
}%
\caption{\label{fig:posterior_BWL}Posterior distributions of all 18 parameters of the BWL* nuclear mass model. The most probable distribution of each parameter of BWL* is shown on top of the respective diagonal plot. The mean values of posterior distributions of all 18 parameters along with the corresponding physical units of BWN* are listed in the fifth column of Table~\ref{tab:mass_models_02}. The optimal parameter set is presented in the sixth column of Table~\ref{tab:mass_models_02}. The deviation of BWL$^\dagger$ and AME2020 binding energies is presented at the nuclear chart with respect to the right color scale. See Fig.~\ref{fig:posterior_BW} for the rest of description.}
\end{figure*}

\begin{table}[t!]
\caption{\label{tab:mass_model_BWL}%
Convergence diagnosis for the reference chain of BWL*. See Sec.~\ref{sec:MultiChain}, Eqs.~(\ref{eq:GelmanRubin}) and (\ref{eq:neff}) for the details.}
\begin{tabular*}{\linewidth}{@{\hspace{10mm}\extracolsep{\fill}}ccc@{\hspace{10mm}}}
\toprule[1.0pt]
\midrule[0.25pt]Parameter & $n_\mathrm{eff}$
 & $\hat{R}$ \\
\midrule[0.25pt]
$\alpha_\mathrm{v}$  & 2505.6 & 1.0003 \\
$\alpha_\mathrm{s}$  & 2503.7 & 1.0012 \\
$\alpha_\mathrm{c}$  & 2512.6 & 1.0030 \\
$c_\mathrm{sym}$     & 2512.5 & 1.0860 \\
$\alpha_\mathrm{xc}$ & 2518.5 & 1.0188 \\
$\alpha_\mathrm{p}$  & 3763.4 & 1.0632 \\
$\alpha_\mathrm{r}$  & 2507.0 & 1.0546 \\
$\alpha_\mathrm{m}$  & 3095.0 & 1.0004 \\
$\beta_\mathrm{m}$   & 2814.3 & 1.0003 \\
$c_\mathrm{m}$       & 3041.4 & 1.0023 \\
$e_\mathrm{m1}$      & 3589.4 & 1.0006 \\
$e_\mathrm{m2}$      & 3102.7 & 1.0020 \\
$k$                  & 2513.9 & 1.0133 \\
$\xi$                & 2631.5 & 1.1757 \\
$\kappa_\mathrm{s}$  & 2681.6 & 1.0462 \\
$g_\mathrm{1}$       & 2627.7 & 1.0034 \\
$g_\mathrm{2}$       & 2606.0 & 1.0020 \\
$\gamma$             & 3062.4 & 1.0027 \\
$\log\sigma$         & 4248.8 & 1.0001 \\

\bottomrule[1.0pt]
\end{tabular*}
\end{table}

Table~\ref{tab:mass_model_BWL} presents the effective sample size, $n_\mathrm{eff}$, and the Gelman-Rubin statistics, $\hat{R}$, for all parameters of the BWL* mass model. Table~\ref{tab:mass_model_BWL} presents the posterior convergence diagnosis for the reference chain of the BWL* model. Most parameters exhibit remarkable convergence, with the potential scale reduction factor $\hat{R}$ close to 1.00 and the effective sample size $n_\mathrm{eff}$ comfortably exceeding 2500. In particular, the newly introduced deformation correction parameters, $g_1, g_2$, and $\gamma$, display sampling efficiencies ($\hat{R} \le 1.0034$), demonstrating that the incorporation of deformation correction successfully regularizes the complex posterior landscape and rectifies the parameter degeneracy observed in BWN*. A minor statistical exception is noted for $\xi$ with $\hat{R} = 1.1757$, indicating a slow convergence of multi Markov chains. The overall diagnosis confirms that the parameter space of the BWL* model is well-converged and robust for posterior inferences despite the slow convergence of multi chain.

Basically, the BWL* model preserves the positively and negatively correlated parameters and major contributing terms of the BWK* and BWN* models. The detailed correlation strengths among the parameters of the BWL* model are presented in the form of the Pearson correlation and non-linear Spearman rank correlation matrices in Table~\ref{tab:bwl_correlation}.

After incorporating the deformation corrections (Eq.~(\ref{eq:BWL})), the condition number of the Pearson correlation matrix for BWN* drops from $1.42 \times 10^6$ to $1.13 \times 10^6$ (BWL*), demonstrating a moderate reduction in parameter degeneracies. Importantly, the condition number of the non-linear Spearman rank correlation matrix for BWL* increases to $1.09 \times 10^3$, compared to $8.14 \times 10^2$ for BWN*, suggesting that using BA-MCMC and Spearman rank correlation analysis method permits us to reveal the reduction of collinearity in parameters of both models. This outcome aligns  with the introduction of more non-linear terms to mass models and points out the limitation of evaluating a mass model with least-squares fit and Pearson correlation matrix.

By incorporating the deformation correction terms with the parameters $ g_1 $, $ g_2 $, and $ \gamma $ (Eq.~(\ref{eq:BWL})), the BWL* model alleviates the multi-modality observed in several posterior distributions of the BWN* model. Such an improvement is indicated by comparing the diagonal plots of $\alpha_\mathrm{v}$, $\alpha_\mathrm{s}$, $\alpha_\mathrm{c}$, and $\alpha_\mathrm{r}$ in Figs.~\ref{fig:posterior_BWN} and \ref{fig:posterior_BWL}. 
The consideration of deformation smooths the likelihood surface, making posterior distributions more unimodal (e.g., covariance plots of $ \alpha_\mathrm{v} $, $ \alpha_\mathrm{s} $, $ \alpha_\mathrm{c} $, and $\alpha_\mathrm{r}$ in Fig.~\ref{fig:posterior_BWL}).

\begin{sidewaystable*}[htbp]
\vspace{85mm}
\caption{\label{tab:bwl_correlation}%
Pearson correlation matrix and Spearman rank correlation matrix of the posterior samples for the BWL* nuclear mass model parameters. See Sec.~\ref{sec:CorrelationDiagnosis}, Eqs.~(\ref{eq:Spearman}) and (\ref{eq:Pearson}) for the details.}
\begin{tabular*}{\linewidth}{@{\hspace{5mm}\extracolsep{\fill}}crrrrrrrrrrrrrrrrrrr@{\hspace{5mm}}}
\toprule[1.0pt]
\midrule[0.25pt]
\multicolumn{20}{c}{Pearson correlation matrix}\\
 & $\alpha_\mathrm{v}$ & $\alpha_\mathrm{s}$ & $\alpha_\mathrm{c}$ & $c_{\rm sym}$ & $\alpha_\mathrm{p}$ & $\alpha_\mathrm{xc}$ & $\alpha_\mathrm{r}$ & $\alpha_\mathrm{m}$ & $\beta_\mathrm{m}$ & $c_\mathrm{m}$ & $e_\mathrm{m1}$ & $e_\mathrm{m2}$ & $k$ & $\xi$ & $\kappa_\mathrm{s}$ & $g_1$ & $g_2$ & $\gamma$ & $\log\sigma$ \\
\midrule[0.25pt]
$\alpha_\mathrm{v}$     & 1.000 & $-0.932$ & $-0.852$ & $-0.369$ & 0.015 & 0.234 & 0.935 & $-0.017$ & $-0.040$ & 0.190 & 0.096 & 0.143 & 0.154 & 0.146 & $-0.257$ & 0.129 & $-0.096$ & $-0.132$ & 0.007 \\
$\alpha_\mathrm{s}$     & $-0.932$ & 1.000 & 0.898 & 0.319 & 0.031 & $-0.536$ & $-0.976$ & 0.022 & 0.013 & $-0.229$ & $-0.090$ & $-0.203$ & $-0.181$ & 0.052 & 0.181 & $-0.074$ & 0.037 & 0.175 & $-0.006$ \\
$\alpha_\mathrm{c}$     & $-0.852$ & 0.898 & 1.000 & 0.030 & $-0.002$ & $-0.630$ & $-0.801$ & 0.002 & $-0.046$ & $-0.048$ & 0.018 & 0.002 & 0.104 & 0.226 & 0.421 & $-0.163$ & 0.151 & 0.116 & $-0.013$ \\
$c_{\rm sym}$  & $-0.369$ & 0.319 & 0.030 & 1.000 & $-0.027$ & 0.156 & $-0.400$ & 0.104 & $-0.167$ & $-0.130$ & $-0.071$ & $-0.052$ & $-0.952$ & $-0.676$ & $-0.667$ & 0.110 & $-0.143$ & 0.221 & 0.016 \\
$\alpha_\mathrm{p}$     & 0.015 & 0.031 & $-0.002$ & $-0.027$ & 1.000 & $-0.063$ & $-0.067$ & 0.059 & $-0.052$ & $-0.008$ & 0.020 & 0.012 & 0.006 & 0.037 & $-0.015$ & 0.023 & $-0.018$ & $-0.009$ & $-0.006$ \\
$\alpha_\mathrm{xc}$  & 0.234 & $-0.536$ & $-0.630$ & 0.156 & $-0.063$ & 1.000 & 0.390 & $-0.009$ & 0.159 & $-0.008$ & $-0.088$ & $-0.004$ & $-0.054$ & $-0.580$ & $-0.090$ & $-0.033$ & 0.030 & $-0.137$ & 0.005 \\
$\alpha_\mathrm{r}$     & 0.935 & $-0.976$ & $-0.801$ & $-0.400$ & $-0.067$ & 0.390 & 1.000 & $-0.036$ & $-0.068$ & 0.283 & 0.144 & 0.279 & 0.250 & 0.076 & $-0.129$ & 0.083 & $-0.031$ & $-0.165$ & 0.003 \\
$\alpha_\mathrm{m}$     & $-0.017$ & 0.022 & 0.002 & 0.104 & 0.059 & $-0.009$ & $-0.036$ & 1.000 & $-0.588$ & $-0.326$ & 0.029 & $-0.001$ & $-0.119$ & $-0.118$ & $-0.109$ & 0.301 & $-0.351$ & 0.114 & $-0.010$ \\
$\beta_\mathrm{m}$      & $-0.040$ & 0.013 & $-0.046$ & $-0.167$ & $-0.052$ & 0.159 & $-0.068$ & $-0.588$ & 1.000 & $-0.206$ & $-0.146$ & $-0.268$ & 0.195 & 0.016 & 0.244 & $-0.590$ & 0.523 & $-0.182$ & 0.017 \\
$c_\mathrm{m}$          & 0.190 & $-0.229$ & $-0.048$ & $-0.130$ & $-0.008$ & $-0.008$ & 0.283 & $-0.326$ & $-0.206$ & 1.000 & 0.198 & 0.416 & 0.136 & 0.136 & 0.036 & $-0.247$ & 0.275 & $-0.221$ & 0.003 \\
$e_\mathrm{m1}$       & 0.096 & $-0.090$ & 0.018 & $-0.071$ & 0.020 & $-0.088$ & 0.144 & 0.029 & $-0.146$ & 0.198 & 1.000 & $-0.274$ & 0.040 & 0.092 & 0.013 & $-0.016$ & 0.074 & $-0.095$ & $-0.021$ \\
$e_\mathrm{m2}$       & 0.143 & $-0.203$ & 0.002 & $-0.052$ & 0.012 & $-0.004$ & 0.279 & $-0.001$ & $-0.268$ & 0.416 & $-0.274$ & 1.000 & 0.035 & $-0.044$ & 0.021 & 0.018 & 0.017 & 0.055 & 0.015 \\
$k$            & 0.154 & $-0.181$ & 0.104 & $-0.952$ & 0.006 & $-0.054$ & 0.250 & $-0.119$ & 0.195 & 0.136 & 0.040 & 0.035 & 1.000 & 0.648 & 0.756 & $-0.171$ & 0.192 & $-0.215$ & $-0.019$ \\
$\xi$          & 0.146 & 0.052 & 0.226 & $-0.676$ & 0.037 & $-0.580$ & 0.076 & $-0.118$ & 0.016 & 0.136 & 0.092 & $-0.044$ & 0.648 & 1.000 & 0.272 & $-0.042$ & 0.055 & $-0.040$ & $-0.008$ \\
$\kappa_\mathrm{s}$     & $-0.257$ & 0.181 & 0.421 & $-0.667$ & $-0.015$ & $-0.090$ & $-0.129$ & $-0.109$ & 0.244 & 0.036 & 0.013 & 0.021 & 0.756 & 0.272 & 1.000 & $-0.291$ & 0.280 & $-0.127$ & $-0.025$ \\
$g_1$          & 0.129 & $-0.074$ & $-0.163$ & 0.110 & 0.023 & $-0.033$ & 0.083 & 0.301 & $-0.590$ & $-0.247$ & $-0.016$ & 0.018 & $-0.171$ & $-0.042$ & $-0.291$ & 1.000 & $-0.928$ & 0.285 & $-0.010$ \\
$g_2$          & $-0.096$ & 0.037 & 0.151 & $-0.143$ & $-0.018$ & 0.030 & $-0.031$ & $-0.351$ & 0.523 & 0.275 & 0.074 & 0.017 & 0.192 & 0.055 & 0.280 & $-0.928$ & 1.000 & $-0.557$ & 0.006 \\
$\gamma$       & $-0.132$ & 0.175 & 0.116 & 0.221 & $-0.009$ & $-0.137$ & $-0.165$ & 0.114 & $-0.182$ & $-0.221$ & $-0.095$ & 0.055 & $-0.215$ & $-0.040$ & $-0.127$ & 0.285 & $-0.557$ & 1.000 & 0.005 \\
$\log\sigma$   & 0.007 & $-0.006$ & $-0.013$ & 0.016 & $-0.006$ & 0.005 & 0.003 & $-0.010$ & 0.017 & 0.003 & $-0.021$ & 0.015 & $-0.019$ & $-0.008$ & $-0.025$ & $-0.010$ & 0.006 & 0.005 & 1.000 \\

\toprule[1.0pt]
\midrule[0.25pt]
\multicolumn{20}{c}{Spearman rank correlation matrix}\\
 & $\alpha_\mathrm{v}$ & $\alpha_\mathrm{s}$ & $\alpha_\mathrm{c}$ & $c_{\rm sym}$ & $\alpha_\mathrm{p}$ & $\alpha_\mathrm{xc}$ & $\alpha_\mathrm{r}$ & $\alpha_\mathrm{m}$ & $\beta_\mathrm{m}$ & $c_\mathrm{m}$ & $e_\mathrm{m1}$ & $e_\mathrm{m2}$ & $k$ & $\xi$ & $\kappa_\mathrm{s}$ & $g_1$ & $g_2$ & $\gamma$ & $\log\sigma$ \\
\midrule[0.25pt]
$\alpha_\mathrm{v}$     & 1.000 & $-0.918$ & $-0.837$ & $-0.403$ & 0.017 & 0.246 & 0.923 & $-0.006$ & $-0.042$ & 0.174 & 0.082 & 0.131 & 0.199 & 0.137 & $-0.256$ & 0.123 & $-0.098$ & $-0.111$ & 0.016 \\
$\alpha_\mathrm{s}$     & $-0.918$ & 1.000 & 0.891 & 0.366 & 0.025 & $-0.550$ & $-0.967$ & 0.009 & 0.008 & $-0.203$ & $-0.074$ & $-0.186$ & $-0.230$ & 0.054 & 0.171 & $-0.061$ & 0.030 & 0.156 & $-0.005$ \\
$\alpha_\mathrm{c}$     & $-0.837$ & 0.891 & 1.000 & 0.088 & 0.003 & $-0.631$ & $-0.775$ & $-0.013$ & $-0.035$ & $-0.036$ & 0.019 & 0.013 & 0.049 & 0.218 & 0.407 & $-0.154$ & 0.144 & 0.104 & $-0.013$ \\
$c_{\rm sym}$  & $-0.403$ & 0.366 & 0.088 & 1.000 & $-0.033$ & 0.098 & $-0.455$ & 0.104 & $-0.165$ & $-0.127$ & $-0.067$ & $-0.050$ & $-0.943$ & $-0.619$ & $-0.585$ & 0.102 & $-0.133$ & 0.213 & 0.017 \\
$\alpha_\mathrm{p}$     & 0.017 & 0.025 & 0.003 & $-0.033$ & 1.000 & $-0.071$ & $-0.055$ & 0.063 & $-0.052$ & $-0.011$ & 0.028 & 0.014 & 0.012 & 0.045 & $-0.024$ & 0.025 & $-0.023$ & 0.005 & 0.000 \\
$\alpha_\mathrm{xc}$  & 0.246 & $-0.550$ & $-0.631$ & 0.098 & $-0.071$ & 1.000 & 0.391 & $-0.006$ & 0.149 & 0.000 & $-0.087$ & 0.008 & $-0.020$ & $-0.562$ & $-0.071$ & $-0.053$ & 0.051 & $-0.130$ & 0.001 \\
$\alpha_\mathrm{r}$     & 0.923 & $-0.967$ & $-0.775$ & $-0.455$ & $-0.055$ & 0.391 & 1.000 & $-0.026$ & $-0.060$ & 0.262 & 0.133 & 0.263 & 0.308 & 0.088 & $-0.111$ & 0.073 & $-0.028$ & $-0.144$ & 0.003 \\
$\alpha_\mathrm{m}$     & $-0.006$ & 0.009 & $-0.013$ & 0.104 & 0.063 & $-0.006$ & $-0.026$ & 1.000 & $-0.578$ & $-0.315$ & 0.029 & $-0.007$ & $-0.123$ & $-0.108$ & $-0.119$ & 0.296 & $-0.339$ & 0.104 & $-0.009$ \\
$\beta_\mathrm{m}$      & $-0.042$ & 0.008 & $-0.035$ & $-0.165$ & $-0.052$ & 0.149 & $-0.060$ & $-0.578$ & 1.000 & $-0.180$ & $-0.140$ & $-0.243$ & 0.194 & 0.001 & 0.246 & $-0.582$ & 0.515 & $-0.178$ & 0.010 \\
$c_\mathrm{m}$          & 0.174 & $-0.203$ & $-0.036$ & $-0.127$ & $-0.011$ & 0.000 & 0.262 & $-0.315$ & $-0.180$ & 1.000 & 0.188 & 0.410 & 0.139 & 0.128 & 0.051 & $-0.249$ & 0.273 & $-0.207$ & 0.004 \\
$e_\mathrm{m1}$       & 0.082 & $-0.074$ & 0.019 & $-0.067$ & 0.028 & $-0.087$ & 0.133 & 0.029 & $-0.140$ & 0.188 & 1.000 & $-0.255$ & 0.034 & 0.084 & 0.003 & $-0.017$ & 0.072 & $-0.090$ & $-0.020$ \\
$e_\mathrm{m2}$       & 0.131 & $-0.186$ & 0.013 & $-0.050$ & 0.014 & 0.008 & 0.263 & $-0.007$ & $-0.243$ & 0.410 & $-0.255$ & 1.000 & 0.033 & $-0.039$ & 0.036 & 0.000 & 0.033 & 0.048 & 0.013 \\
$k$            & 0.199 & $-0.230$ & 0.049 & $-0.943$ & 0.012 & $-0.020$ & 0.308 & $-0.123$ & 0.194 & 0.139 & 0.034 & 0.033 & 1.000 & 0.611 & 0.678 & $-0.167$ & 0.185 & $-0.205$ & $-0.016$ \\
$\xi$          & 0.137 & 0.054 & 0.218 & $-0.619$ & 0.045 & $-0.562$ & 0.088 & $-0.108$ & 0.001 & 0.128 & 0.084 & $-0.039$ & 0.611 & 1.000 & 0.196 & $-0.019$ & 0.029 & $-0.028$ & $-0.009$ \\
$\kappa_\mathrm{s}$     & $-0.256$ & 0.171 & 0.407 & $-0.585$ & $-0.024$ & $-0.071$ & $-0.111$ & $-0.119$ & 0.246 & 0.051 & 0.003 & 0.036 & 0.678 & 0.196 & 1.000 & $-0.300$ & 0.287 & $-0.119$ & $-0.025$ \\
$g_1$          & 0.123 & $-0.061$ & $-0.154$ & 0.102 & 0.025 & $-0.053$ & 0.073 & 0.296 & $-0.582$ & $-0.249$ & $-0.017$ & 0.000 & $-0.167$ & $-0.019$ & $-0.300$ & 1.000 & $-0.926$ & 0.289 & $-0.006$ \\
$g_2$          & $-0.098$ & 0.030 & 0.144 & $-0.133$ & $-0.023$ & 0.051 & $-0.028$ & $-0.339$ & 0.515 & 0.273 & 0.072 & 0.033 & 0.185 & 0.029 & 0.287 & $-0.926$ & 1.000 & $-0.546$ & 0.000 \\
$\gamma$       & $-0.111$ & 0.156 & 0.104 & 0.213 & 0.005 & $-0.130$ & $-0.144$ & 0.104 & $-0.178$ & $-0.207$ & $-0.090$ & 0.048 & $-0.205$ & $-0.028$ & $-0.119$ & 0.289 & $-0.546$ & 1.000 & 0.008 \\
$\log\sigma$   & 0.016 & $-0.005$ & $-0.013$ & 0.017 & 0.000 & 0.001 & 0.003 & $-0.009$ & 0.010 & 0.004 & $-0.020$ & 0.013 & $-0.016$ & $-0.009$ & $-0.025$ & $-0.006$ & 0.000 & 0.008 & 1.000 \\
\bottomrule[1.0pt]
\end{tabular*}
\end{sidewaystable*}

\begin{table*}
\caption{\label{tab:mass_models_01} Nuclear mass models according to the BW and BWK formulations.}
\begin{tabular*}{\linewidth}{@{\hspace{1.0mm}\extracolsep{\fill}}rrrrrrr@{\hspace{1.0mm}}}
\toprule[1.0pt]
\midrule[0.25pt]
   &  BW\footnotemark[1]  &  BW*\footnotemark[2]  & BW$^{\dagger}$\footnotemark[3]  &  BWK\footnotemark[1]  & BWK*\footnotemark[2]  &  BWK$^{\dagger}$\footnotemark[3]  \\ 
\midrule[0.25pt]
$\alpha_\mathrm{v}$ (MeV)    &  $15.5255\pm0.0242$   &  $15.7110\pm0.0273$   & $15.7186$   &  $16.490\pm0.0659$   & $16.5157\pm0.0285$  &  $16.5453$  \\ 
$\alpha_\mathrm{s}$ (MeV)    &  $-16.8949\pm0.0753$  &  $-17.5030\pm 0.0855$ & $-17.5267$ &  $-25.5618\pm0.4630$   & $-25.6311\pm0.1907$ &  $-25.8050$  \\ 
$\alpha_\mathrm{c}$ (MeV)    &  $-0.7022\pm0.0017$   &  $-0.7138\pm0.0019$   & $ -0.7143$  &  $-0.7614\pm0.0031$   & $-0.7608\pm0.0016$  &  $-0.7611$  \\ 
$\alpha_\mathrm{sym}$ (MeV)  &  $-22.9874\pm0.0603$  &  $-23.4116\pm0.0678$  & $-23.4309$ &  $-32.5777\pm0.3042$  & $-32.8138\pm0.1629$                  &  $-33.1780$   \\ 
$\alpha_\mathrm{p}$ (MeV)    &  --                   &  --                   & --  &  $11.0409\pm0.4608$         & $10.9671\pm0.4145$   &  $10.7263$  \\ 
$\alpha_\mathrm{xc}$ (MeV)   &  --                   &  --                   & --   &  $1.6997\pm0.0743$         & $1.6347\pm0.0693$   &  $1.6144$  \\ 
$\alpha_\mathrm{w}$ (MeV)    &  --                   &  --                   & -- &  $-61.7229\pm2.8712$        & $-62.4516\pm1.9926$                &  $ -65.3746$   \\ 
$\alpha_\mathrm{st}$ (MeV)   &  --                   &  --                   & --  &  $61.1172\pm1.5853$         & $62.3074\pm0.9459$                  &  $64.3523$   \\ 
$\alpha_\mathrm{r}$ (MeV)    &  --                   &  --                   & --  &  $13.3315\pm0.7883$         & $13.5406\pm0.3210$  &  $13.9222$  \\ 
$\alpha_\mathrm{m}$ (MeV)    &  --                   &  --                   & --  &  $-2.0293\pm0.0419$         & $-1.9808\pm0.0422$  &  $-1.9796$  \\ 
$\beta_\mathrm{m}$ (MeV)     &  --                   &  --                   & --   &  $0.1595\pm0.0045$         & $0.1538\pm0.0047$   &  $0.1541$  \\ 
$c_\mathrm{m}$ (MeV)         &  --                   &  --                   & --                  &  --          & --  &  --  \\ 
$e_\mathrm{m1}$ (MeV)        &  --                   &  --                   & --                  &  --          & --   &  --  \\ 
$e_\mathrm{m2}$              &  --                   &  --                   & --                  &  --          & --   &  --  \\ 
$k$                          &  --                   &  --                   & --                  &  --          & --   &  --  \\ 
$\xi$                        &  --                   &  --                   & --                  &  --          & --   &  --  \\ 
$\kappa_\mathrm{s}$                   &  --                   &  --                   & --                  &  --          & --   &  --  \\ 
$g_1$                        &  --                   &  --                   & --                  &  --          & --                  &  --   \\ 
$g_2$                        &  --                   &  --                   & --                  &  --          & --                  &  --   \\ 
$\gamma$                     &  --                   &  --                   & --                  &  --          & --                  &  --   \\ 
log$\sigma$                  &  --                   &   $1.0992\pm0.0151$                   & $1.0869$                  &   --          & $0.4707\pm  0.0151$                &   $0.470$   \\
$\sigma(B)(\text{MeV})$\footnotemark[4]&  --    &  $3.002\pm0.045$              & $2.965$             &  --    & $1.601\pm0.024$             &  1.600  \\
RMS($B$)(MeV) & $3.194$ & -- & $ 3.0003$ & $1.625$ & -- & $1.5984$ \\
\bottomrule[1.0pt]
\end{tabular*}
\footnotetext[1]{\fontsize{7pt}{7pt}\selectfont  Parameters obtained by \citet{Wu2025} using the least-squares fitting method.}
\footnotetext[2]{\fontsize{7pt}{7pt}\selectfont  Mean values of posterior distribution deduced in this work using Bayesian analysis with MCMC sampling.}
\footnotetext[3]{\fontsize{7pt}{7pt}\selectfont  Optimal parameters obtained by calculating RMS over all MCMC samples.}
\footnotetext[4]{\fontsize{7pt}{7pt}\selectfont  $\sigma(B)$ is from Eq.~(\ref{ref:sigma(b)}).}
\end{table*}

\begin{table*}
\caption{\label{tab:mass_models_02} Nuclear mass models according to the BWN and BWL formulations.}
\begin{tabular*}{\linewidth}{@{\hspace{1.0mm}\extracolsep{\fill}}rrrrrrrr@{\hspace{1.0mm}}}
\toprule[1.0pt]
\midrule[0.25pt]
   &  BWN\footnotemark[1]  &  BWN*\footnotemark[2]  & BWN$^{\dagger}$\footnotemark[3]    & BWL*\footnotemark[2]  &  BWL$^{\dagger}$\footnotemark[3]  &  BWL$^{\ddagger}$\footnotemark[5]\\ 
\midrule[0.25pt]
$\alpha_\mathrm{v}$ (MeV)    &  $16.7043\pm0.0398$   &  $16.7532\pm0.0385$   & $16.7879$    & $16.7527\pm0.0329$  &  $16.7753$   & $16.7429$ \\ 
$\alpha_\mathrm{s}$ (MeV)    &  $-26.3000\pm0.2824$  &  $-26.6693\pm 0.2771$ & $-26.9145$   & $-26.4845\pm0.2331$ &  $-26.6043$  & $ -26.3482$ \\ 
$\alpha_\mathrm{c}$ (MeV)    &  $-0.7615\pm0.0020$   &  $-0.7632\pm0.0020$   & $ -0.7647$   & $-0.7647\pm0.0017$  &  $-0.7656$   & $-0.7637$ \\ 
$c_\mathrm{sym}$ (MeV)       &  $-35.3636\pm0.2648$  & $-35.2349\pm0.2033$   & $-35.2807$   & $-35.7408\pm0.2246$ &  $-35.8330$  & $-35.9206$ \\ 
$\alpha_\mathrm{p}$ (MeV)    &  $5.9751\pm0.1303$    &  $5.8966\pm0.1190$    & $5.8727$     & $5.9391\pm0.1133$   &  $5.9649$    & $5.9310$ \\ 
$\alpha_\mathrm{xc}$ (MeV)   &  $1.4405\pm0.0473$    &  $1.4506\pm0.0557$    & $1.4673$     & $1.4224\pm0.0463$   &  $1.4194$    & $1.3870$ \\ 
$\alpha_\mathrm{w}$ (MeV)    &  --                   &  --                   & --           & --                  &  --          & -- \\ 
$\alpha_\mathrm{st}$ (MeV)   &  --                   &  --                   & --           & --                  &  --          & -- \\ 
$\alpha_\mathrm{r}$ (MeV)    &  $14.1287\pm0.4749$   &  $14.8315\pm0.4522$   & $15.2511$    & $14.4738\pm0.3877$  &  $14.6690$   & $14.2559$ \\ 
$\alpha_\mathrm{m}$ (MeV)    &  $-1.0877\pm0.0277$   &  $-1.0747\pm0.0274$   & $-1.0786$    & $-0.7776\pm0.0286$  & $-0.7982$    & $-0.7686$ \\ 
$\beta_\mathrm{m}$ (MeV)     &  $0.1615\pm0.0028$    &  $0.1567\pm0.0030$    & $0.1578$     & $0.1134\pm0.0037$   &  $0.1161$    & $0.1120$  \\ 
$c_\mathrm{m}$ (MeV)         &  $-0.2343\pm0.0076$   &  $-0.2243\pm0.0080$   & $-0.2241$    & $-0.2651\pm0.0078$  &  $-0.2623$   & $-0.2651$ \\ 
$e_\mathrm{m1}$ (MeV)        &  $5.4713\pm0.1229$    &  $5.4954\pm0.1129$    & $5.5262$     & $5.3122\pm0.1160$   &  $5.2897$    & $5.2660$  \\ 
$e_\mathrm{m2}$              &  $-0.0444\pm0.0020$   &  $-0.0420\pm0.0018$   & $-0.0410$    & $-0.0452\pm0.0020$  &  $-0.0450$   & $-0.0457$ \\ 
$k$                          &  $2.0829\pm0.0288$    &  $2.0638\pm0.0224$    & $2.0630$     & $2.1193\pm0.0236$   &  $2.1243$    & $2.1398$ \\ 
$\xi$                        &  $1.2216\pm0.0313$    &  $1.2090\pm0.0295$    & $1.2073$     & $1.2598\pm0.0250$   &  $1.2694$    & $1.2877$ \\ 
$\kappa_\mathrm{s}$          &  $0.2491\pm0.0314$    &  $0.1807\pm0.0281$    & $0.1710$     & $0.1614\pm0.0272$   &  $0.1611$    & $0.1807$ \\ 
$g_1$                        &  --                   &  --                   & --           & $0.0232\pm0.0010$   &  $0.0226$    & $ 0.0235$\\ 
$g_2$                        &  --                   &  --                   & --           & $-0.7184\pm0.0291$  &  $ -0.6987$  & $ -0.7265$\\ 
$\gamma$                     &  --                   &  --                   & --           & $3.5907\pm0.4412$   &  $3.2989$    & $3.5508$  \\ 
log$\sigma$                  &  --                   &  $-0.1430\pm0.0151$   & $-0.1526$    & $-0.2644\pm  0.0148$&  $-0.2837$   & $-0.2642$ \\
$\sigma(B)(\text{MeV})$\footnotemark[4] &  --        &  $0.867\pm0.013$      & $0.858$      & $0.768\pm0.011$     &  $ 0.753$    & $0.768$   \\
RMS($B$)(MeV)                & $0.887$               & --                    & $0.8633$     & --                  & $0.7583$     & $0.7635$  \\
\bottomrule[1.0pt]
\end{tabular*}
\footnotetext[1]{\fontsize{7pt}{7pt}\selectfont Parameters obtained by \citet{Wu2025} using the least-squares fitting method.}
\footnotetext[2]{\fontsize{7pt}{7pt}\selectfont Mean values of posterior distribution deduced in this work using Bayesian analysis with MCMC sampling.}
\footnotetext[3]{\fontsize{7pt}{7pt}\selectfont Optimal parameters obtained by comparing RMS over all MCMC samples.}
\footnotetext[4]{\fontsize{7pt}{7pt}\selectfont $\sigma(B)$ is from Eq.~(\ref{ref:sigma(b)}).}
\footnotetext[5]{\fontsize{7pt}{7pt}\selectfont Parameters based on BA-MCMC and the WS4 masses \cite{WS4database}.}
\end{table*}

The posterior distributions of the BWL* model reveal new non-linear yet monotonic correlations among $g_1$, $g_2$, and $\gamma$ parameters (Eq.~(\ref{eq:deformation_coeff})). In Table~\ref{tab:bwl_correlation}, the compensation between the deformation correction coefficients, $g_1$ and $g_2$, is indicated by \textbf{$\rho(g_1,g_2)=-0.926$}. These two parameters calibrate the deformation correction factor (Eq.~(\ref{eq:deformation_fac})) in such a way that $g_1$ strengthens the first-term deformation correction, while $g_2$ weakens the second-term correction, smoothening the overall deformation correction strength applied on the macroscopic energy.

The Spearman rank correlation matrices of $\rho(g_1,\gamma)$ and $\rho(g_2,\gamma)$ are 0.285 and $-0.546$, respectively. The positive (negative) value of $\rho(g_1,\gamma)$ ($\rho(g_2,\gamma)$) indicates that $\gamma$ is positively (negatively) correlated with $\rho(g_1,\gamma)$ ($\rho(g_2,\gamma)$). See Eqs.~(\ref{eq:deformation_fac}) and (\ref{eq:deformation_quench}) for details. These $g_1$, $g_2$, and $\gamma$ factors are optimized through BA-MCMC to describe the total binding energy of strongly deformed nuclei. The Spearman rank correlation matrix of $\rho(\beta_\mathrm{m},g_2)=0.515$ quantifies the positive correlation between the quadratic shell correction and the second term of deformation correction in Eq.~(\ref{eq:deformation_coeff}). The Spearman rank correlation matrices of $\rho(\alpha_\mathrm{m},g_1)=0.296$ and $\rho(\alpha_\mathrm{m},g_2)=-0.339$ suggest a positive correlation of $\alpha_\mathrm{m}$-$g_1$ and a negative correlation of $\alpha_\mathrm{m}$-$g_2$, respectively.

Here we discuss some cases to show the correlations among $g_1$, $g_2$, $\gamma$, and $\alpha_\mathrm{m}$. A first-order expansion of the macroscopic deformation factor in Eq.~(\ref{eq:deformation_fac}) after substituting Eq.~(\ref{eq:deformation_coeff}) gives the deformation contributions proportional to $g_1$ and $g_2$, 
\begin{align}
\Delta B_{g_1}^{(1)} &= B^\mathrm{macro} g_1 A^{1/3}
(\beta_2^2+2\beta_4^2+3\beta_6^2) \, , \nonumber\\
\Delta B_{g_2}^{(1)} &= B^\mathrm{macro} g_2 A^{-1/3}
(\beta_2^2+4\beta_4^2+9\beta_6^2) \, ,
\end{align}
of which non-leading-order terms are negligible. For nuclei with $|\beta_2|\!<\!0.01$ and $P\!=\!{\nu_\mathrm{p} \nu_\mathrm{n}}/({\nu_\mathrm{p} + \nu_\mathrm{n}})\!=\!0 $, the total contribution of $\Delta B_{g_1}^{(1)}$ and $ \Delta B_{g_2}^{(1)}$ is actually smaller than $0.01$~MeV. For instance, for $^{90}$Zr, $\Delta B_{g_1}^{(1)}$ and $\Delta B_{g_2}^{(1)}$ are 0.004~MeV and $-0.006$~MeV, respectively, and for $^{208}$Pb, $\Delta B_{g_1}^{(1)}$ and $\Delta B_{g_2}^{(1)}$ are +0.003~MeV and $-0.004$~MeV, respectively. The macroscopic energies of $^{90}$Zr and $^{208}$Pb are quenched by the deformation parameter $|\beta_2|$ in the deformation correction, and $B_i^{\text{shell}}$ consisting of the deformation quenching factor (Eq.~(\ref{eq:deformation_quench})) and shell correction (Eq.~(\ref{eq:BWL_shell})) is dominant in their binding energies.

For $^{238}$U, $^{240}$Pu, and $^{244}$Cm, their $\beta_2$ are 0.2155, 0.2213, and 0.2261, respectively \cite{WS4database}. Meanwhile, the corresponding $\Delta B_{g_1}^{(1)}$ and $\Delta B_{g_2}^{(1)}$ for $^{238}$U, $^{240}$Pu, and $^{244}$Cm in units of MeV are 14.056 and $-13.171$, 14.439 and $-13.128$, and 14.394 and $-12.352$, respectively. These $\Delta B_{g_1}^{(1)}$ and $\Delta B_{g_2}^{(1)}$ with $g_1$ and $g_2$ form a counterbalance relation, almost canceling each other in contributing to the macroscopic LDM energy. Following the complete product in Eq.~(\ref{eq:deformation_fac}), the net macroscopic deformation corrections for $^{238}$U, $^{240}$Pu, and $^{244}$Cm are $+0.883$, $+1.310$, and $+2.041$~MeV, respectively. The suppressive role of the deformation quenching factor (Eq.~(\ref{eq:deformation_quench})) is important to calibrate the dominant $\alpha_\mathrm{m}P$ term in contributing the bound energy to $B_i^{\text{shell}}$ of $^{238}$U, $^{240}$Pu, and $^{244}$Cm. The quenched $B_i^{\text{shell}}$ in the BWL$^\dagger$ model for $^{238}$U, $^{240}$Pu, and $^{244}$Cm are $-4.498$, $-5.031$, and $-5.730$~MeV, respectively. Combining the impact of the deformation correction factor (Eq.~(\ref{eq:deformation_fac})) and deformation quenching factor (Eq.~(\ref{eq:deformation_quench})), the residuals of $^{238}$U, $^{240}$Pu, and $^{244}$Cm of BWL$^\dagger$ are reduced from $+0.641$ to $+0.058$~MeV, from $+0.776$ to $+0.141$~MeV, and from $+0.795$ to $+0.065$~MeV, respectively, comparing with those of BWN$^\dagger$.

Collectively, these correlations demonstrate that the deformation correction terms ($g_1$, $g_2$, $\gamma$) and the shell correction terms ($\alpha_\mathrm{m}$, $\beta_\mathrm{m}$) in the BWL* model exhibit significant physical compensation relationships. In addition, the introduction of the deformation parameters $g_1$, $g_2$, and $\gamma$ regularizes the overall posterior distribution becoming more symmetric, reducing (high) parameter degeneracy in BWK* and BWN*.

The compensation relations between $k$-$\xi$, $k$-$\kappa_\mathrm{s}$, and $\xi$-$\kappa_\mathrm{s}$ become more pronounced in the covariance plots of the BWL* model (Fig.~\ref{fig:posterior_BWL}). This indicates that the inclusion of deformation corrections leads to a set of less degenerate posterior distributions. Consequently, the four parameters $ c_{\rm sym} $, $k$, $\xi$, and $ \kappa_\mathrm{s} $ supports the synergy of the isospin-dependent symmetry energy terms from the WS4 model in the BWN* formulation by Wu et al. (2025).

\begin{figure*}[htbp]
\includegraphics[width=0.5\linewidth]{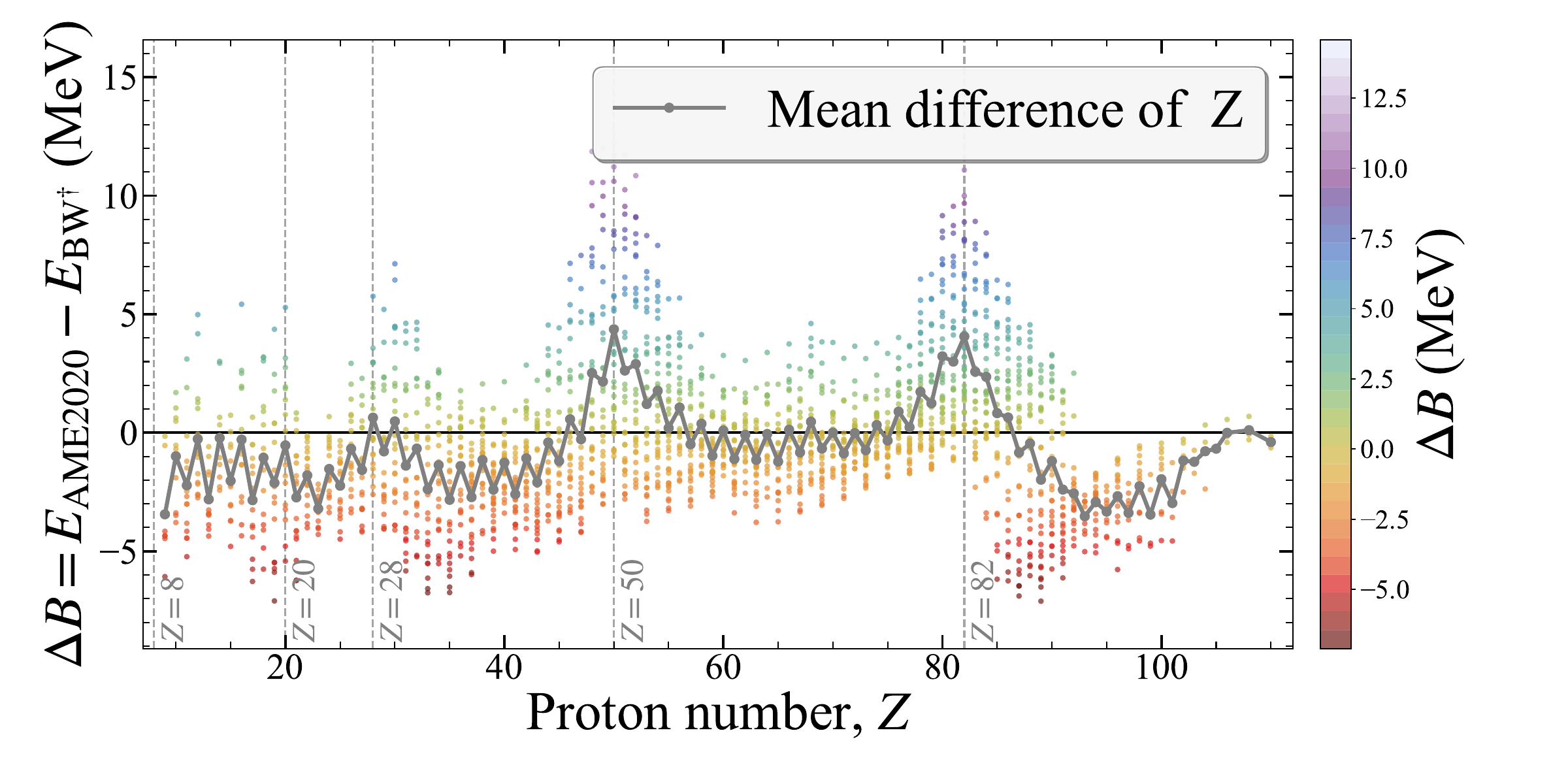}%
\includegraphics[width=0.5\linewidth]{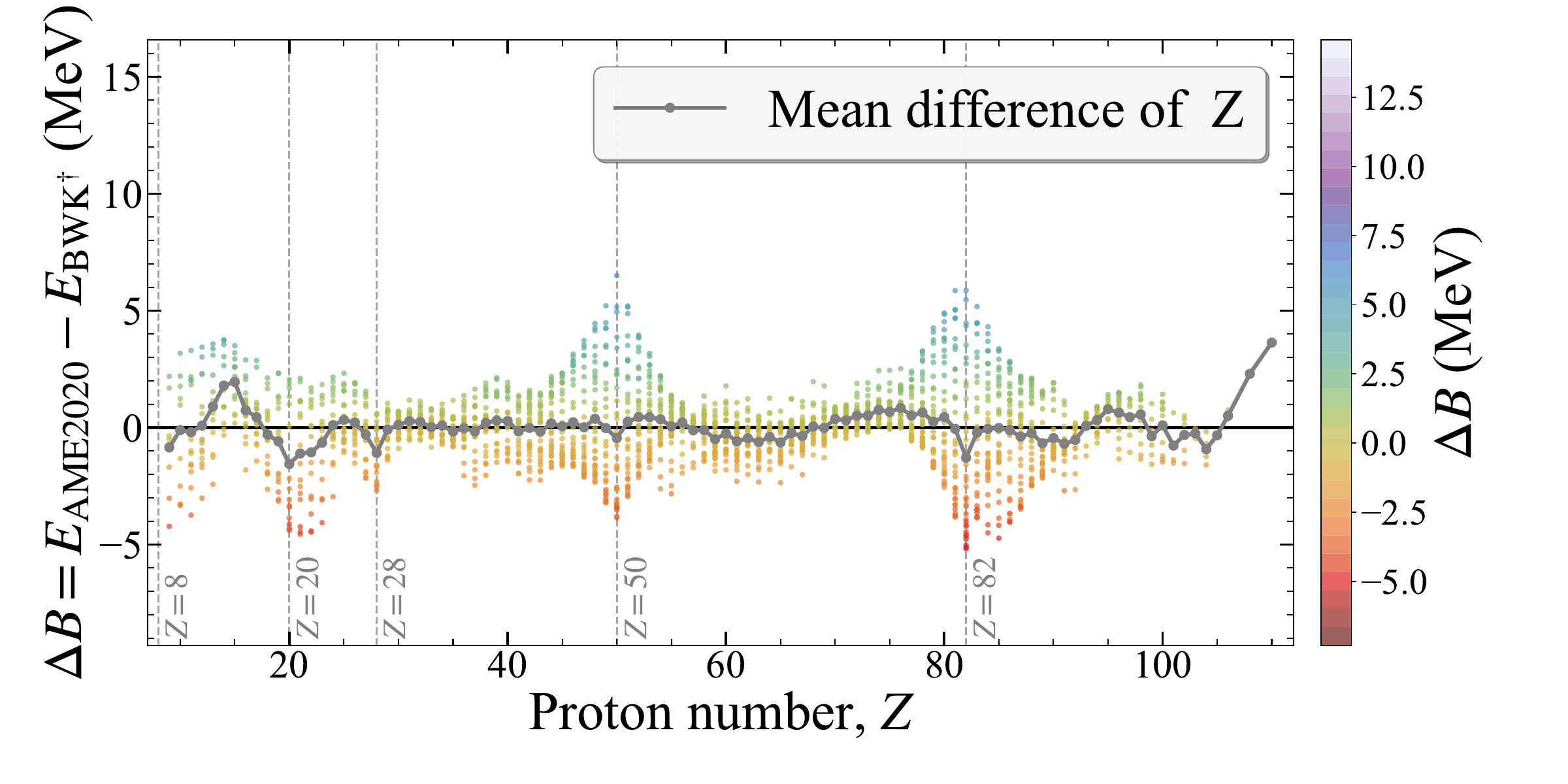}
\includegraphics[width=0.5\linewidth]{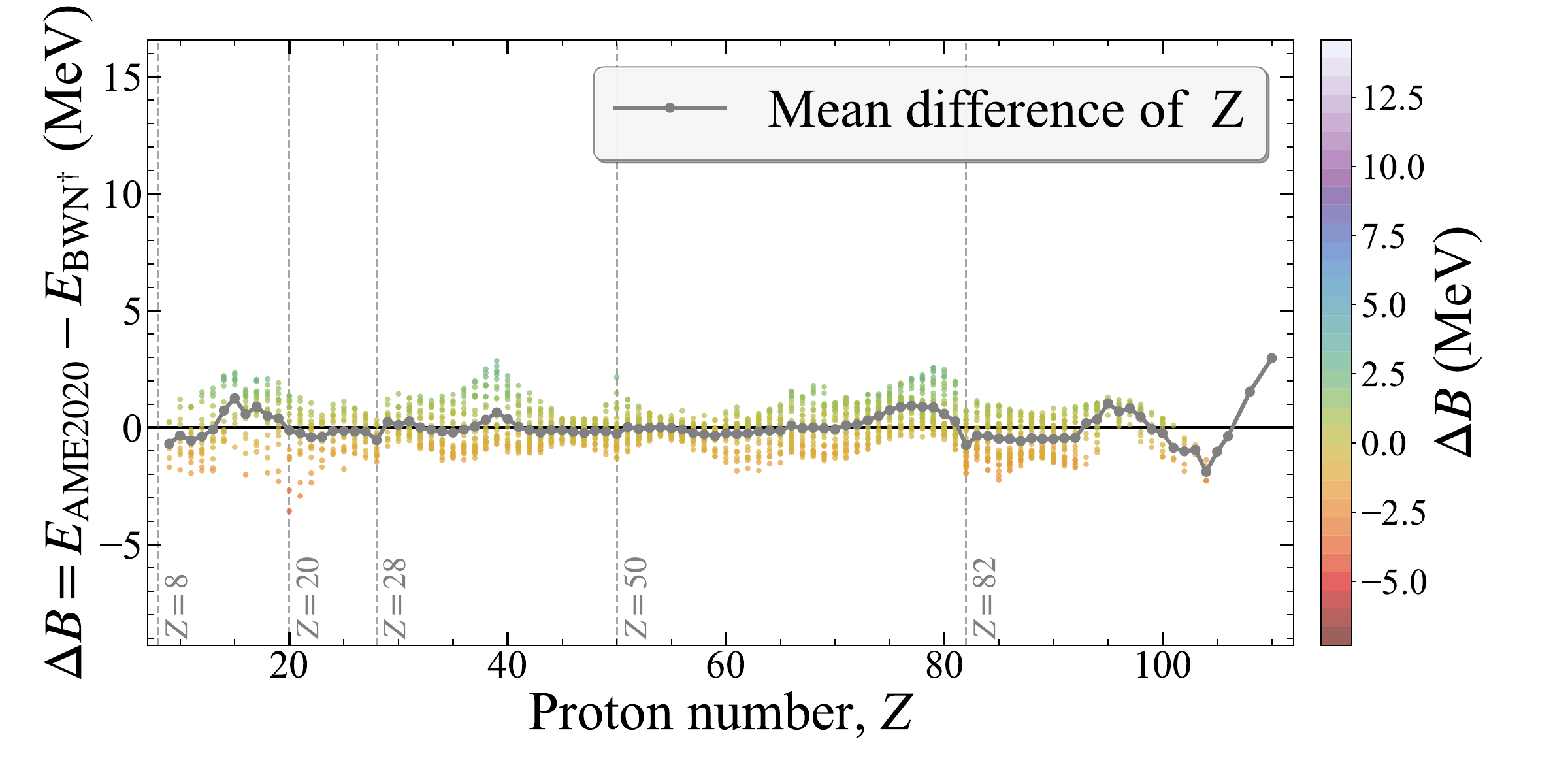}%
\includegraphics[width=0.5\linewidth]{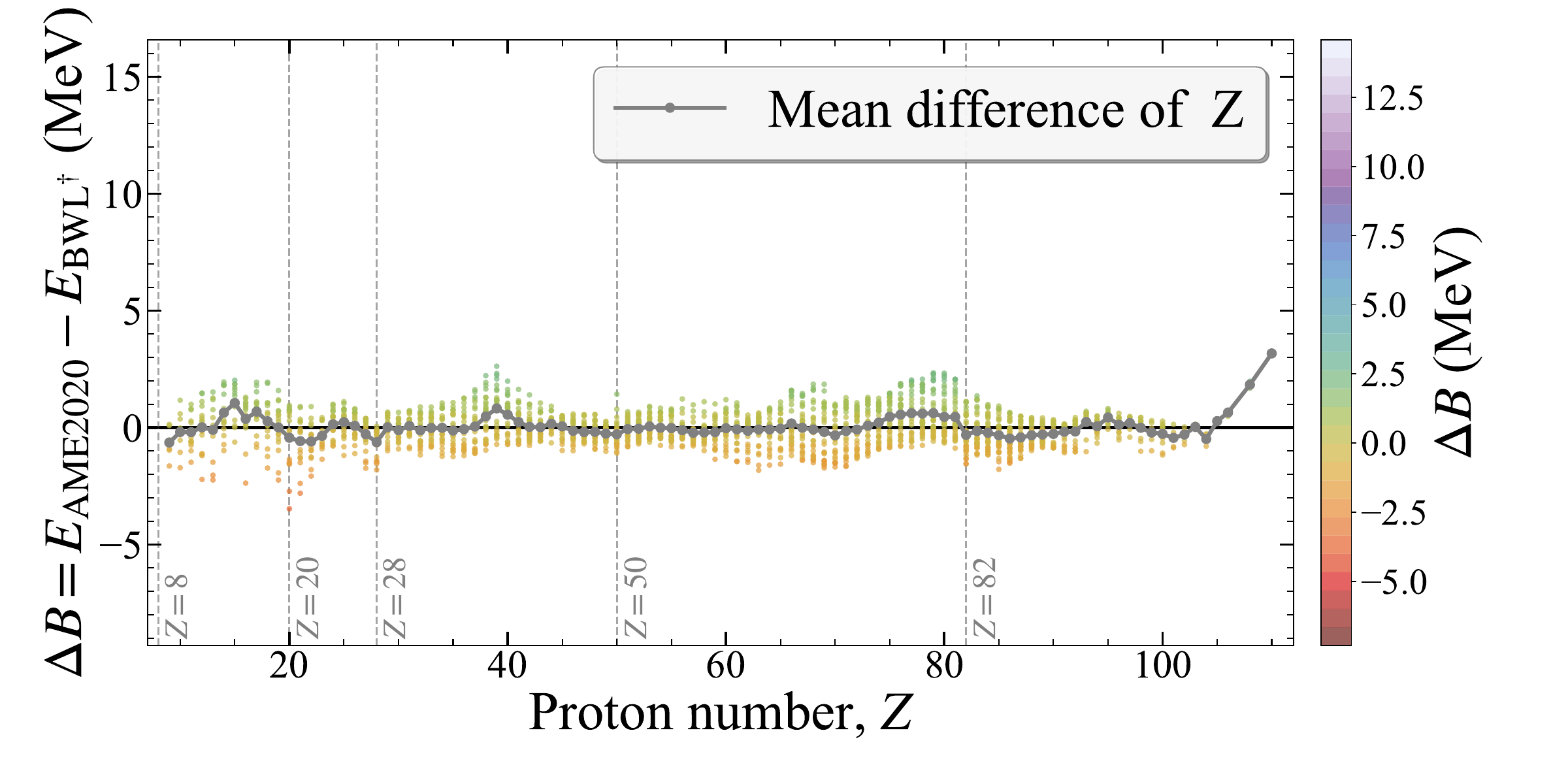}
\vspace{-8mm}
\caption{\label{fig:BW_mean}Deviations of nuclear binding energies between AME2020 and BW$^\dagger$, BWK$^\dagger$, BWN$^\dagger$, BWL$^\dagger$.}
\end{figure*}

\subsection{General improvement imposed by deformation}
\label{sec:general_improvement}

By comparing the mean difference of $Z$ between the BWN$^\dagger$ and BWL$^\dagger$ models, the introduction of the macro deformation correction factor (Eq.~(\ref{eq:deformation_fac})) noticeably reduces the overall residuals between the theoretical and AME2020 mass values for nuclei of $ Z > 82 $ (Fig.~\ref{fig:BW_mean}). The most pronounced improvement is observed in the actinide region of $Z=89$-103. The mean absolute residual error (MAE) is reduced from $0.7517$~MeV to $0.3566$~MeV, a reduction of approximately 53~\%. Nuclei in this open-shell actinide region are generally well deformed, and their equilibrium shapes reflect the interplay between macroscopic deformation energy and shell correction \cite{Nilsson2005,Brack1972,Moller2016}. For the light nucleus region of $ Z < 20 $, the MAE is reduced from $1.0653$~MeV to $0.8710$~MeV, a reduction of $\approx\!18~\%$, particularly for the nuclei of $ Z < 15 $. Moreover, near the proton magic number $ Z = 50 $, the residual of BWL$^\dagger$ is up to 713~keV lower than that of BWN$^\dagger$. This demonstrates that the deformation quenching factor (Eq.~(\ref{eq:deformation_quench})) suppresses shell effects in both single-magic nuclei and open-shell nuclei. In fact, this improvement is also benefited from the more comprehensive characterization of shell effects achieved by the BWN$^\dagger$ model.

\begin{figure*}[htbp]
\includegraphics[width=1.0\linewidth]{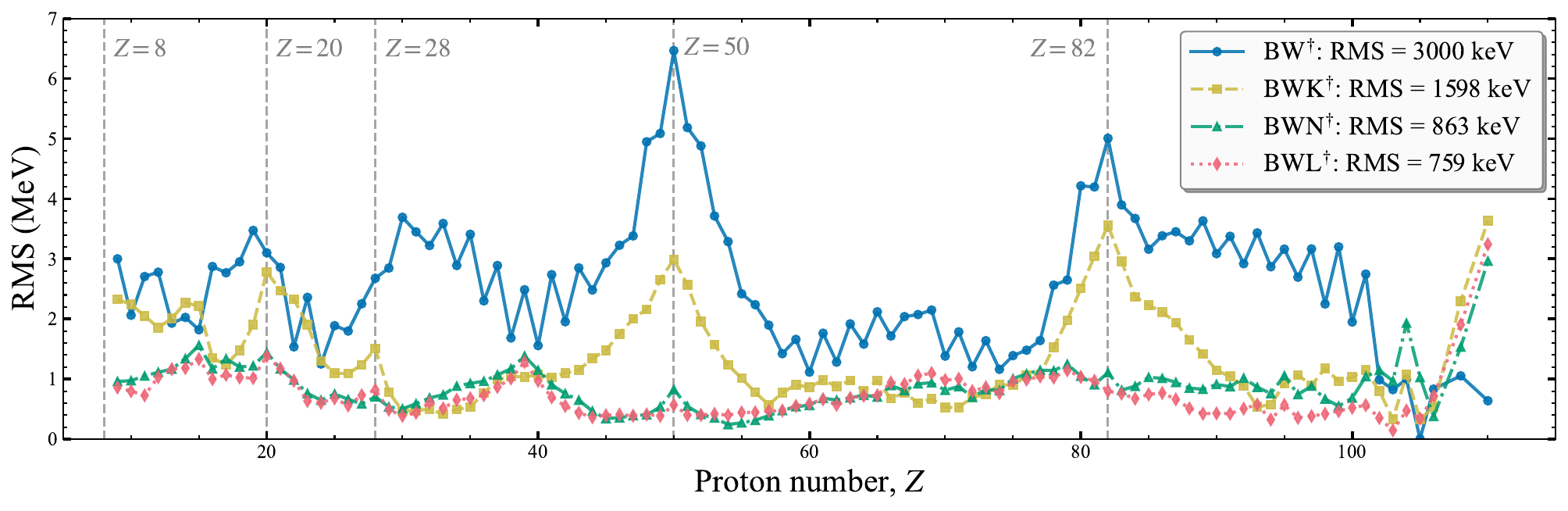}%
\vspace{-5mm}
\caption{\label{fig:rms_four_models}Root mean square deviations (RMSs) of comparing experimental and theoretical binding energies from BW$^\dagger$, BWK$^\dagger$, BWN$^\dagger$, and BWL$^\dagger$ for each isotopic chain. See Tables~\ref{tab:mass_models_01} and \ref{tab:mass_models_02} for BA-MCMC optimized mass models.}
\end{figure*}

The BWL$^\dagger$ model reduces the global RMS deviation of the BWN$^\dagger$ model from 0.863~MeV to 0.759~MeV for the 2242 selected nuclei, corresponding to an improvement of approximately 12~\%. As shown in Fig.~\ref{fig:rms_four_models}, among the mass models, BWL$^\dagger$ produces the lowest average residuals for the light-nuclei region of $Z<20$ but the residual of each nuclide is still not significantly reduced, suggesting that the descriptive capability of the deformation corrections in BWL$^\dagger$ should have been exhausted, while the shell correction terms can still be improved.

For the medium-heavy mass region of $Z\approx 28$-50, the BWL$^\dagger$ model produces the lowest overall RMS among all mass models. Nevertheless, the deformation is weak for the region around $Z=40$, e.g., the average quadrupole deformation \(\overline{\beta_2}=0.0962\) for yttrium isotopes ($Z=39$), whereas the shell effect dominates and causes these isotopes to become more overbound than the experimental counterparts. For the rare-earth region, the results of mean residuals (Fig.~\ref{fig:BW_mean}) and RMS (Fig.~\ref{fig:rms_four_models}) also demonstrate that the shell correction term (Eq.~(\ref{eq:BWL_shell})) can be further improved.

For the actinide and superheavy region of $Z>82$, especially near $Z\!\approx\!100$, the mean residuals and RMS of BWL$^\dagger$ remain superior to other models. The improvement in predicting unmeasured masses is of considerable importance for estimating fission behavior and syntheses of superheavy nuclei \cite{Pei2009,Oganessian2015}.

For historical large discrepancies along the isotopic chains of magic proton numbers $Z=50$, and 82, the RMS of these chains are successively reduced with refinements and BA-MCMC optimization on these models. Both BWN$^\dagger$ and BWL$^\dagger$ reasonably reproduce experimental data. The difference in RMS between these two models is only up to 308~keV, indicating that the spherical shell effect inherited from BWN$^\dagger$ to BWL$^\dagger$ helps reproduce experimental data.

\begin{figure}[t]
\centering
\includegraphics[width=\columnwidth]{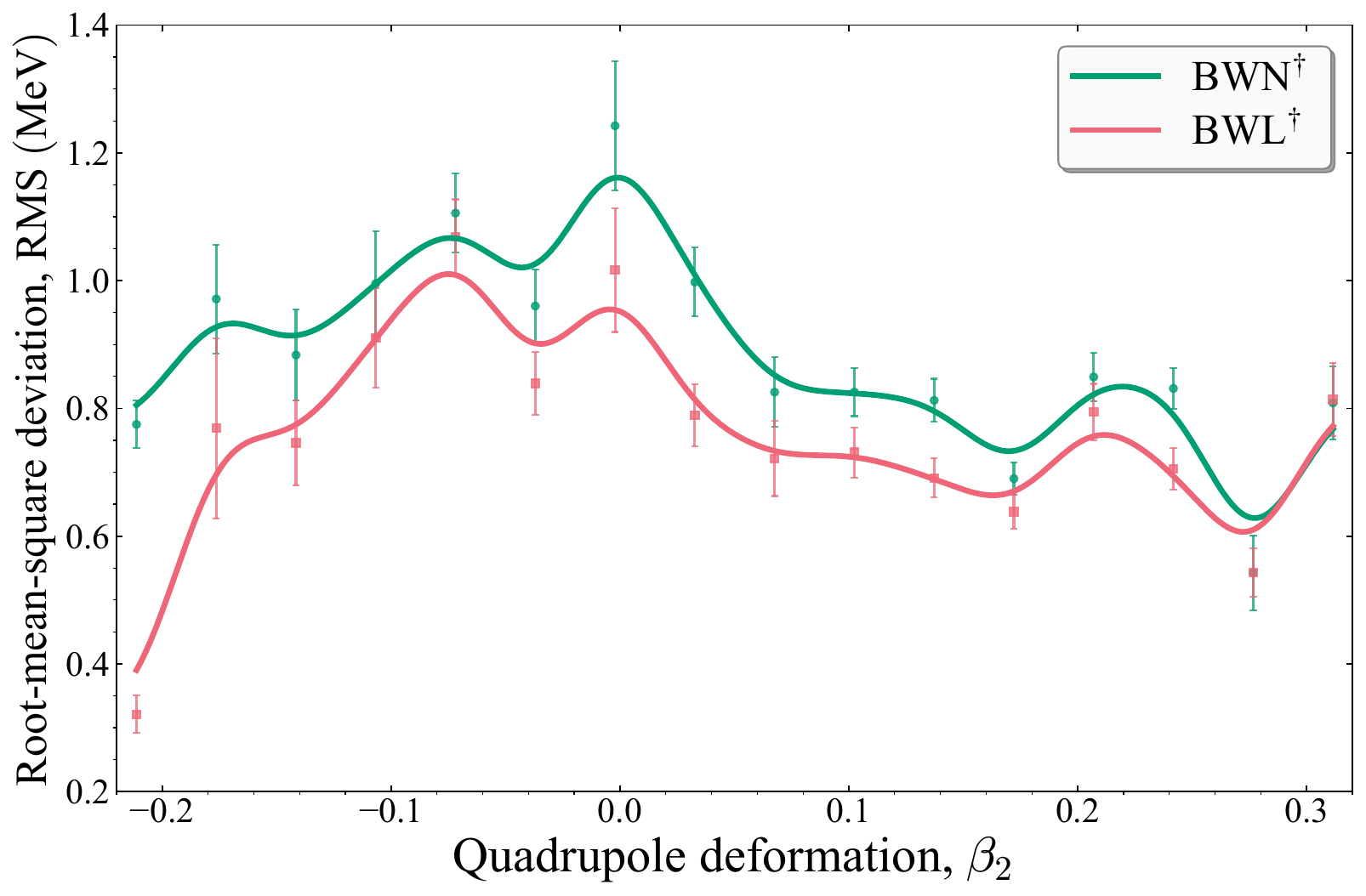}
\caption{\label{fig:rms_beta2}RMS deviation as a function of the quadrupole deformation parameter $\beta_2$. Both BWL$^\dagger$ (red line) and BWN$^\dagger$ (blue line) are plotted with equal width bins over the interval of $-0.2635\leq\beta_2\leq0.3289$, and bin width, $\Delta\beta_2\approx0.035$. As deformation is not included in BWN$^\dagger$, RMS deviations of BWN$^\dagger$ are merely plotted for comparison with following the same interval as nuclei of BWL$^\dagger$. Bins containing fewer than 8 nuclei are discarded, leaving 16 effective points as shown.}
\end{figure}

We plot the RMS deviation as a function of the WS4 quadrupole deformation parameter $\beta_2$ in Fig.~\ref{fig:rms_beta2} to quantify the deformation dependence of the BWN$^\dagger$ and BWL$^\dagger$ models. BWL$^\dagger$ yields lower RMS values than those of BWN$^\dagger$ as shown by the 14 effective points at $\beta_2\lessapprox0.25$, with a count-weighted average RMS reduction of about $0.105$~MeV. The reduction is about $0.09$--$0.13$~MeV in the region $\beta_2\approx0.10$--$0.24$, where many 
deformed nuclei are located. The two largest $\beta_2$ bins, containing 73 and 23 nuclei, show nearly identical RMS values, i.e., 0.24-5~keV. The analysis indicates that the inclusion of deformation-dependent terms in BWL improve the mass description over a broad range of deformed nuclei rather than only through a global refit of parameters.

\subsection{Advantage of Bayesian Analysis with\\Adaptive Metropolis-Hastings\\Markov Chain Monte Carlo Sampling}
\label{sec:advantage}

BA-MCMC offers full posterior distributions of all model parameters, of which the posterior distributions naturally incorporate physically motivated priors and jointly infer the noise level \cite{Saito2024}. The least-squares fit method can only provide point estimates and asymptotic errors, however. The advantage of BA-MCMC enables us to directly analyze the joint distributions and strong correlations among parameters from the posterior distribution plots, thereby supplying a statistical evidence (reliable basis) for further physical simplification of the model.

For instance, the posterior distributions of the BW* model manifest a more symmetrical shape compared to the overall distributions of the BWK* model, which somehow undergo a break of the symmetrical distributions (Figs.~\ref{fig:posterior_BW} and \ref{fig:posterior_BWK}). The posterior distributions for each parameter of the BWK* model slightly deviate from Gaussian shapes and show skewness with longer tails, shifting the overall joint-posterior distributions from an elliptical shape. The underlying reason is due to the introduction of Coulomb exchange, Wigner, pairing, and surface-symmetry energies and implementation of physical boundary constraints on the BWK* parameters (e.g., $\beta_\mathrm{m} > 0$), when combined with Gaussian priors, produce non-Gaussian posteriors \cite{Lundquist2020}. Although posterior distributions of the BWK* model are asymmetric, the BWK* model offers a substantially more comprehensive foundation for the subsequent development of the BWN* and BWL* models.

Further posterior analysis of the BWK model reveals that the Wigner term is positively correlated with the symmetry energy term (the posterior plot of $\alpha_\mathrm{sym}$-$\alpha_\mathrm{w}$ in Fig.~\ref{fig:posterior_BWK}, the Spearman rank correlation matrix of $\rho(\alpha_\mathrm{sym},\alpha_\mathrm{w})=0.660$ in Table~\ref{tab:bwk_correlation}), while both the symmetry energy term and the Wigner term exhibit negative correlations with the surface symmetry term (the posterior plots of $\alpha_\mathrm{sym}$-$\alpha_\mathrm{st}$ and $\alpha_\mathrm{w}$-$\alpha_\mathrm{st}$ in Fig.~\ref{fig:posterior_BWK}, the Spearman rank correlation matrices of 
$\rho(\alpha_\mathrm{sym},\alpha_\mathrm{st})=-0.886$
and
$\rho(\alpha_\mathrm{w},\alpha_\mathrm{st})=-0.632$ 
in Table~\ref{tab:bwk_correlation}). The posterior plots of these terms show pronounced correlations that are highly similar in physical meaning and possess compensatory effects, underpinning the merge of these highly correlated terms in the development of BWN model. Thereby constructing a more concise nuclear mass model with preserving the essential physical meaning. BA-MCMC not only validates the rationality of this combination, but also quantifies the uncertainty of the merged parameters through the full posterior distributions, providing important theoretical guidance and statistical analysis for the future new optimized macroscopic-microscopic nuclear mass models.

The posterior structure of the parameter space of BWN* model (Fig.~\ref{fig:posterior_BWN}) is significantly altered in the BWL* model (Fig.~\ref{fig:posterior_BWL}) by introducing a deformation correction factor in the macroscopic part and a deformation quenching factor in the shell correction term (Eq.~(\ref{eq:BWL})) \cite{Spanier1988, Wang2013WS3, Wang2014}. The multi-modality observed in several BWN* parameters (i.e., $ \alpha_\mathrm{v} $, $ \alpha_\mathrm{s} $, $ \alpha_\mathrm{c} $, $ \alpha_\mathrm{r} $, and $ \alpha_\mathrm{r} $) is alleviated in the BWL* model. The general posterior distributions become more unimodal, symmetric, and smoothened. Analysis of posterior distributions of the BWN* and BWL* models reveals that the newly introduced deformation parameters $ g_1 $ and $ g_2 $ effectively smooth the likelihood surface, reduce parameter degeneracies, and regulate the joint-posterior distributions. 

To verify the generality of the deformation scheme (quadrupole and high-multipole deformation and quenched shell correction), we applied exactly the same deformation form to the BWK$^\dagger$ model. The consideration of deformation further reduces the RMS deviation of the BWK$^\dagger$ model from 1.598~MeV to 1.476~MeV, demonstrating the universality of this deformation scheme in phenomenological models. Moreover, with replacing AME2020 by the WS4 mass data \cite{WS4database} as the counterpart pseudo-data, we reoptimize BWL with BA-MCMC and obtain a set of BWL$^\ddagger$ parameters (the seventh column of Table~\ref{tab:mass_models_02}) with an RMS of $ 0.763 $~MeV. The difference between BWL* and BWL$^\ddagger$ parameters is merely up to 9.3~\%. In fact, the BWL$^\ddagger$ parameters are within the uncertainty of BWL*, implying the universality of BA-MCMC in optimizing BWL for reproducing AME2020 and WS4 masses, of which the RMS of comparing AME2020 and WS4 masses is 0.283~MeV \cite{Wang2014}.

Using the BA-MCMC method, the BWN model is optimized and produces an RMS of 863~keV. Meanwhile, the introduction of deformation correction terms to the BWN model reduces the RMS of the original BWN model from 863~keV to 759~keV (Table~\ref{tab:mass_models_02}), which corresponds to an approximate reduction in RMS of 12~\%. Moreover, the MAE drops by $\approx\!53$~\% in the actinide region ($ Z > 82 $) and by $\approx\!18$~\% in the light-nuclei region ($ Z < 20 $), while the average residual as a function of $ Z $ decreases noticeably. These results demonstrate that the implicit incorporation of deformation not only optimizes the posterior estimates of the parameters but also enhances the model's descriptive power in strongly deformed and light-mass regions, providing a solid foundation for the future development of more refined macroscopic-microscopic nuclear mass models \cite{Barbero2012}.

Nevertheless, all BW$^\dagger$, BWK$^\dagger$, and BWN$^\dagger$ mass models produce a systematic deviation from AME2020 in the rare-earth region ($  A \approx 150  $-$  190  $, $  Z \approx 60  $-$  76  $) with local RMS of 1.678~MeV, 0.720~MeV, and 0.814~MeV, respectively. Although the BWL$^\dagger$ model considers deformation, it produces a local RMS of 0.840~keV, little higher than that of BWN$^\dagger$ (Figs.~\ref{fig:BW_mean} and \ref{fig:rms_four_models}). According to \citet{Moller2016}, 82.3~\%~($\beta_2 > 0.10$) of nuclei in the rare-earth region exhibit prominent quadrupole deformation of the shape of a prolate ellipsoid \cite{Nilsson2005}. The shell correction in the BWL$^\dagger$ model, which is dominant by the valence proton-neutron interaction factor, $P ={\nu_\mathrm{p} \nu_\mathrm{n}}/(\nu_\mathrm{p} + \nu_\mathrm{n})$, and the piecewise constant, $ \delta_{\rm shell} $, even when damped by the deformation quenching factor $ \exp(-\gamma \sum_k \beta_k^2) $, does not describe deformation-induced shell gaps well, which require a microscopic description of single-particle spectra \cite{Brack1972}. Therefore, the remaining residuals in the rare-earth region should be due to the limitation of the present phenomenological shell correction rather than the introduced deformation correction. As shown in Fig.~\ref{fig:BW_mean}, the BWL$^\dagger$ residuals still exhibit mean deviations similar to those of BWN$^\dagger$ in this region.

The posterior space in the BA-MCMC analysis increases as the number of physical terms and parameters increases from 5 (for the BW model) to 19 (for the BWL model). 
We observe that the high-dimensional posterior space causes a slower chain mixing, and thus the Gelman-Rubin \(\hat{R}\) values could exceed 1.1 for some parameters, resulting in non-Gaussian or multi-modal posterior distributions. Therefore, longer MCMC chains (40 million steps per model in this work), much longer burn-in periods (the first 20 million samples used for exploring a parameter space have to be discarded), and higher resolution of covariance adaptation have to be implemented to generate reliable posteriors. 
The cost-effectiveness of the MCMC exploration becomes stressed. The computational speed is reduced by more than 93~\% in the same hardware configuration by comparing the analysis for the BW and BWL models. More sophisticated GPU with adequate GPU memory resources may help overcome such performance bottlenecks or delays.

\section{Summary and Conclusions}
\label{sec:summary}

In this work, we have systematically investigated the nuclear mass models of the Bethe-Weizs{\"a}cker variants, i.e., BW, BWK, BWN, and BWL, using full Bayesian analysis with Markov chain Monte Carlo sampling (BA-MCMC).

BA-MCMC reveals rich compensation (degeneracy) relations among model parameters. For the BWK$^\dagger$ model, strong negative correlations among the volume, surface, Coulomb, and curvature terms, and between the linear $\alpha_\mathrm{m}$ and quadratic shell correction coefficients $\beta_\mathrm{m}$. The BWN$^\dagger$ model preserves most of the non-symmetric correlations after introducing the isospin-dependent symmetry energy, while new correlations appear among the symmetry parameters \(c_{\rm sym}\), \(k\), \(\xi\), and \(\kappa_\mathrm{s}\). By implicitly incorporating deformation corrections, the BWL$^\dagger$ model reshapes the posterior parameter space and alleviates the multi-modality and strong degeneracies observed in BWN$^\dagger$, resulting in more unimodal and symmetric posterior distributions.

Quantitatively, the BWL model further refines the BWN model by introducing the deformation scheme and reduces the global RMS deviation for 2242 nuclei from 863~keV to 759~keV, corresponding to an improvement of $\approx\!12$~\%. In particular, in strongly deformed regions (the rare-earth region of $Z \approx 60$-76 and the actinide region of $ Z > 82 $), the BWL model not only significantly reduces the mean residual but also substantially decreases the error amplitude, with the mean absolute error in the actinide region reduced by $\approx\!53$~\%, providing higher precision training database for developing AI mass models and interpretable physics features for the physics-informed AI approaches, estimating the masses of unmeasured neutron-rich nuclei important for studying (fissions of) superheavy nuclei and island of stability, and for investigating nucleosynthesis along rapid-proton (rapid-neutron) capture process.

We show that using the Spearman rank correlation matrix is more appropriate than the Pearson correlation coefficient for analyzing the non-linear correlations among parameters of a mass model.

The current study indicates that BA-MCMC is a robust tool for diagnosing parameter degeneracies and guiding the refinement of presently available (and future) nuclear mass models. The new mass model, BWL, with its clearer physical picture and more regular parameter space, provides a solid foundation for further improvements on macroscopic-microscopic nuclear mass models. Future work may introduce more refined deformation-dependent shell corrections and region-dependent pairing treatments to achieve cost-effective and higher-precision phenomenological nuclear mass predictions.

\bigskip
\begin{acknowledgments}
This work was supported by the National Natural Science Foundation of China (No. 11775277) and the Science Foundation of Zhejiang Sci-Tech University (No. 25062123-Y). We greatly appreciate the computing resource provided by the Yukawa Institute Computer Facility, i.e., Yukawa-21 and Heian at the Yukawa Institute of Theoretical Physics (YITP) of Kyoto University, Japan. 
\end{acknowledgments}

\newpage
\noindent
{\bf Declaration of Generative AI and AI-assisted technologies in the writing process}
\medskip\\
During the preparation of this manuscript, the authors only used the default but capability-limited \textsc{WriteFull} package provided by Overleaf to check word spellings and did not use any artificial intelligence tools to improve word choice, readability, and clarity. 

\bigskip
\noindent
{\bf Data Availability Statement}
\smallskip\\
Data in Figs.~\ref{fig:posterior_BW}, \ref{fig:posterior_BWK}, \ref{fig:posterior_BWN}, and \ref{fig:posterior_BWL} are publicly available at the Zenodo repository \cite{Zenodo}. These data can also be used with \href{https://numpy.org}{\textsc{NumPy}} and \href{https://corner.readthedocs.io}{\textsc{Corner}} by \citet{corner} to produce these figures. 

\bigskip
\noindent
{\bf CRediT authorship contribution statement}
\medskip\\
\href{https://orcid.org/0009-0003-1508-8447}{\bf Xiangnan Lee{\includegraphics[height=8pt]{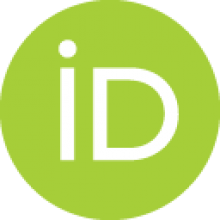}}}: Conceptualization of this study, Data curation, Formal analysis, Investigation, Methodology, Software, Validation, Visualization, Writing - original draft preparation, Writing - review and edition. 
\href{https://orcid.org/0000-0001-6646-0745}{\bf Yi~Hua~Lam{\includegraphics[height=8pt]{Orcid-ID.png}}}: Conceptualization of this study, Formal analysis, Funding acquisition, Investigation, Methodology, Project administration, Resources, Software, Supervision, Validation, Writing - original draft preparation, Writing - review and edition.
\href{http://orcid.org/0009-0006-9932-9616}{\bf Zi-Ao Zhang{\includegraphics[height=8pt]{Orcid-ID.png}}}: Data curation, Software. 
\href{http://orcid.org/0009-0009-8362-8795}{\bf Jayke Ren{\includegraphics[height=8pt]{Orcid-ID.png}}}: Data curation, Software, Visualization.



\bibliography{NuclMass_MCMC} 

\end{document}